\documentclass[]{TEAI}
\usepackage{helvet}

\usepackage{amsmath} 
\usepackage{natbib}
\usepackage{graphicx}
\usepackage{subcaption} 
\usepackage{wrapfig}

\usepackage[toc,page,header]{appendix}
\usepackage[utf8]{inputenc} % allow utf-8 input
\usepackage[T1]{fontenc}    % use 8-bit T1 fonts
\usepackage{hyperref}       % hyperlinks
\usepackage{url}            % simple URL typesetting
\usepackage{subfiles}
\usepackage{multicol}
\usepackage{booktabs}       % professional-quality tables
\usepackage{lmodern}        % scalable Latin Modern fonts
\usepackage{amsfonts}       % blackboard math symbols
\usepackage{nicefrac}       % compact symbols for 1/2, etc.
\usepackage{microtype}      % microtypography

\usepackage{amssymb} 
\usepackage{fontawesome} 
\usepackage{url} 

\usepackage{titletoc}

\usepackage{tikz}  
\usepackage{comment} 
\usepackage{tabularx}  
\usepackage{booktabs}  
\usepackage{minitoc}

\usepackage{booktabs}
\usepackage{array}
\usepackage{etoolbox}

\definecolor{lightblue}{RGB}{200, 230, 255}  
\definecolor{headerblue}{RGB}{150, 200, 255} 

\usepackage{pgfplots}
\usepackage[utf8]{inputenc} % allow utf-8 input
\usepackage[T1]{fontenc}    % use 8-bit T1 fonts
\usepackage{hyperref}       % hyperlinks
\usepackage{url}            % simple URL typesetting
\usepackage{booktabs}       % professional-quality tables
\usepackage{amsfonts}       % blackboard math symbols
\usepackage{nicefrac}       % compact symbols for 1/2, etc.
\usepackage{microtype}      % microtypography
\usepackage{graphicx}
\usepackage{float}
\usepackage{comment}
\usepackage{multirow} % For multi-row cells
\usepackage{amsmath} % For \text command if needed inside math mode\Delta
\usepackage{makecell} % For multi-line cells and better vertical spacing in cells
\usepackage{siunitx}  % For better number alignment (optional but recommended)
\usepackage{pgf-pie} % Package for creating pie charts
\usepackage{subcaption}
\usepackage{wrapfig}
\usepackage{algpseudocode}
\usepackage{algorithm}

\usepackage{bbding}
\usepackage[export]{adjustbox}
\newcommand{\modelname}{SPEED}

\usepackage{ragged2e}      % for \RaggedRight in tabularx
\usepackage{tabularx}       % For tables with fixed total width and auto-adjusting columns
\usepackage{array}          % For advanced column formatting (like >{\centering\arraybackslash}X)
\usepackage{caption}        % Recommended for figures/tables, but we'll do simple text below images here.
\usepackage{enumitem}
\usepackage{pifont}
\usepackage[hang,flushmargin]{footmisc} 

\usepackage{tcolorbox}

\usepackage{tcolorbox}
\tcbuselibrary{breakable}
\tcbuselibrary{skins}
\usepackage{tabularx}
\usepackage{listings}

\usepackage{xcolor}
\definecolor{codegreen}{rgb}{0,0.6,0}
\definecolor{codegray}{rgb}{0.5,0.5,0.5}
\definecolor{codepurple}{rgb}{0.58,0,0.82}
\definecolor{backcolour}{rgb}{0.95,0.95,0.92}
\definecolor{mediumtealblue}{rgb}{0.0, 0.33, 0.71}
\definecolor{darkpastelgreen}{rgb}{0.01, 0.75, 0.24}
\definecolor{azure}{rgb}{0.0, 0.5, 1.0}

\definecolor{myadd}{RGB}{255,160,122}
\definecolor{myreduce}{RGB}{100,149,237}
\definecolor{myblue}{RGB}{66,106,179}
\definecolor{myred}{HTML}{b22c46}
\definecolor{myyellow}{HTML}{decb00}

\usepackage{tcolorbox}
\definecolor{lightblue}{HTML}{18282e}
\definecolor{lighterblue}{HTML}{f2fafd} 
\newtcolorbox{abox}{colback=lighterblue,colframe=lightblue, width=\textwidth}

\title{{\fontsize{15pt}{22pt}\selectfont SPEED: One-Step Pixel Diffusion for High-quality \\ Video Frame Interpolation}}

\author{
{\normalfont\large\mdseries
\begin{tabular}{@{}c@{}}
Zihao Zhang\textsuperscript{1,2,$\star$} \quad
Haoyu Zhao\textsuperscript{1,$\star$} \quad
Siqian Yang\textsuperscript{2} \quad
Yidi Wu\textsuperscript{2} \quad
 \\[0.1em]
Yudong Jiang\textsuperscript{2}
\quad
Zuxuan Wu\textsuperscript{1,$\dagger$}
\end{tabular}
}
}

\affiliation[1]{\mbox{Fudan University}}
\affiliation[2]{\mbox{Bilibili Inc}}

\abstract{
\begin{abstract}

Despite the success of diffusion models in Video Frame Interpolation (VFI), existing methods still suffer from two critical limitations.
First, latent diffusion inevitably loses fine-grained details when reconstructing images from latent representations back to the pixel space. Second, multi-step sampling incurs prohibitive memory consumption and inference latency. To address these issues, we propose \modelname, a one-step pixel diffusion framework for high-quality VFI. Specifically, \modelname~employs a progressive multi-stage architecture with dynamic patch scaling to effectively learn multi-scale motion, structural, and appearance representations. Furthermore, we propose a novel Noise-Update-Only Attention mechanism to prevent semantic degradation of the clean condition frames while reducing the computational overhead by nearly 50\%. Besides, we introduce a Drift-aware Timestep Sampling strategy coupled with a tailored training objective to directly predict images in the pixel space, enabling one-step inference without compromising the quality of the generated frames.
Extensive experiments show that \modelname\ achieves state-of-the-art performance. On SNU-FILM, SPEED reduces LPIPS by 8.8\% while delivering 63.3\% faster inference and 10.6\% lower memory usage. On challenging 4K benchmarks, it further surpasses prior methods by up to 51.5\% in LPIPS.\looseness=-1
\end{abstract}

}

\correspondence{Zuxuan Wu at \email{zxwu@fudan.edu.cn}}
\checkdata[Accepted by]{\href{https://2026.acmmm.org/}{ACM International Conference on Multimedia (ACMMM), 2026.}}
\checkdata[Project Page]{\href{https://bbldcver.github.io/SPEED/}{https://bbldcver.github.io/SPEED/}}

\begin{document}
\maketitle
\renewcommand{\thefootnote}{}
\footnotetext{$^\star$Equal contribution.\\$^\dagger$Corresponding authors.}
\renewcommand{\thefootnote}{\arabic{footnote}}

\vspace{-1.5em}

\section{Introduction}

Video Frame Interpolation (VFI) is a fundamental task in computer vision and multimedia processing, aimed at synthesizing intermediate frames between given starting and ending frames. It plays a crucial role in numerous practical applications, including slow-motion generation~\cite{safa, slomo, videoinr}, frame rate up-conversion~\cite{RIFE, EMA, IFRNet}, novel view synthesis~\cite{LearningTo, blur, deepstereo}, and video compression~\cite{vidcom, NCM}.
Recently, propelled by the rapid advancement of generative models~\cite{ddpm, ldm, dit, sd3, qwen-image, wan, dynamictrl, cameranoise}, the latest VFI methods~\cite{eden,tlbvfi,LDMVFI,LBBDM,VIDIM,hifi} have increasingly adopted the diffusion paradigm. By formulating VFI as a conditional generation task, these methods successfully generate intermediate content for complex, non-linear motions that traditionally pose significant challenges to optical flow-based methods~\cite{EMA, stmfnet, RIFE, SGMVFI, vfimamba, lcmamba, bimvfi}.

\begin{figure}
  \centering
  \begin{tikzpicture}
        \node[anchor=south west, inner sep=0] (image) at (0,0) {
            \includegraphics[width=1.0\linewidth]{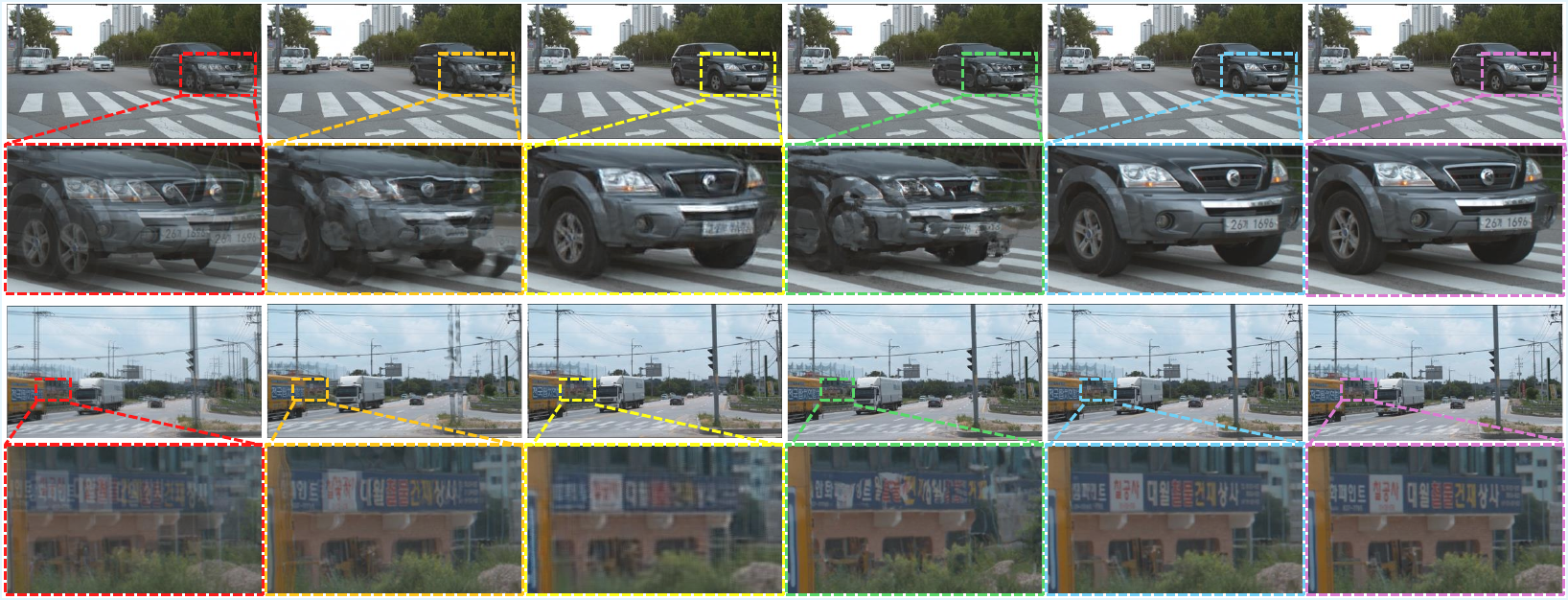}
        };
        \node[anchor=north west, color=black] at (0.6,6.8) {\fontsize{10}{10}\selectfont \textbf{Overlay}};
        \node[anchor=north west, color=black] at (3.1,6.8) {\fontsize{10}{10}\selectfont \textbf{LDMVFI}~\cite{LDMVFI}};
        \node[anchor=north west, color=black] at (6.0,6.8) {\fontsize{10}{10}\selectfont \textbf{EDEN}~\cite{eden}};
        \node[anchor=north west, color=black] at (8.6,6.8) {\fontsize{10}{10}\selectfont 
        \textbf{TLB-VFI}~\cite{tlbvfi}};
        \node[anchor=north west, color=black] at (11.2,6.8) {\fontsize{10}{10}\selectfont \textbf{SPEED (Ours)}};
        \node[anchor=north west, color=black] at (14.8,6.8) {\fontsize{10}{10}\selectfont \textbf{GT}};
    \end{tikzpicture}
  \caption{Visual comparison of SPEED with state-of-the-art latent diffusion VFI models on XTest4K~\cite{XVFI}. The leftmost image is the overlaid image of the starting and ending frames. By performing VFI entirely in pixel space, SPEED eliminates the detail loss caused by latent reconstruction. This enables our model to preserve and synthesize sharp, high-quality details (\textit{e.g.}, text, edges, and micro-textures) that are blurred or distorted by existing methods.}
  \label{fig:teaser}
\end{figure}
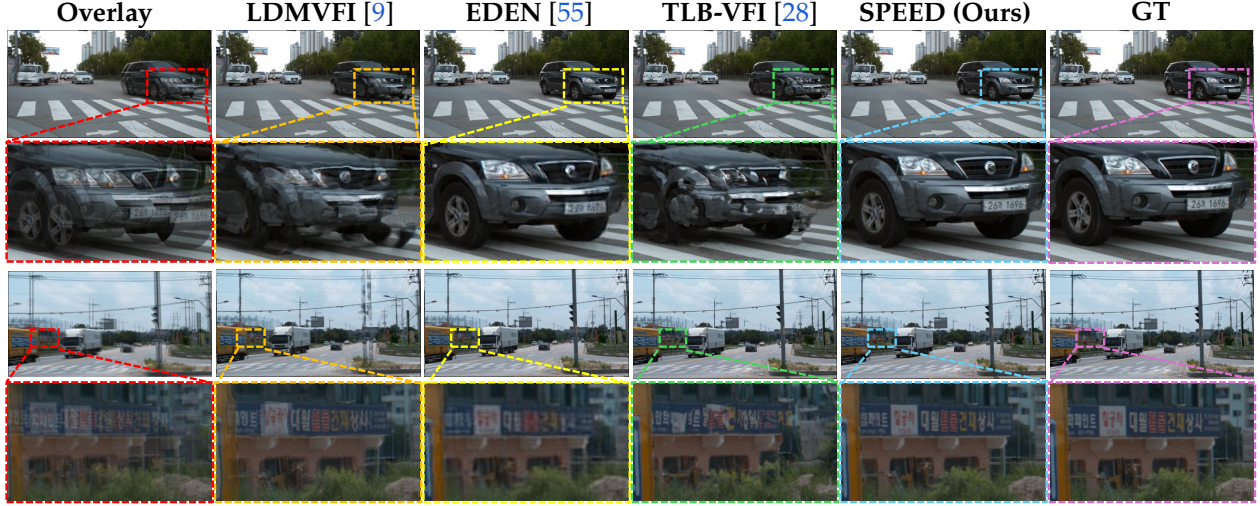

However, existing diffusion-based VFI methods~\cite{LDMVFI, LBBDM, eden, tlbvfi} are predominantly Latent Diffusion Models (LDMs)~\cite{ldm}, which perform the denoising process within a compressed latent space encoded by a pre-trained Variational Autoencoder (VAE)~\cite{vae}. Despite their well-established advantages over optical flow in complex motion scenarios, these latent diffusion methods suffer from two critical bottlenecks:
1) \textbf{Information Loss in Encoding and Decoding:} Compressing the natural pixel space into a low-dimensional latent representation inevitably loses vital fine-grained details. This imposes a hard upper bound on reconstruction quality, causing the decoded outputs to frequently exhibit blurred patterns and missing fine details in previous methods, as shown in Fig.~\ref{fig:teaser}.
2) \textbf{Inference Inefficiency:} Although multiple sampling in diffusion models enables diverse plausible outputs, it also incurs significant memory overhead and computational latency. For instance, generating a single $720 \times 1280$ interpolated frame requires 3678.7 ms using LDMVFI~\cite{LDMVFI} and consumes 19.06 GB of memory with CBBD~\cite{LBBDM}. These massive resource requirements strictly prohibit their practical deployment for higher-quality generation and high-resolution (e.g., 4K) scenarios.

To address these limitations, we propose SPEED, a novel framework for high-quality intermediate frame generation that operates entirely in the pixel space and performs one-step denoising. Although pixel diffusion naturally avoids the VAE-induced loss of fine details, training directly in pixel space remains challenging due to the difficulty of jointly modeling large-scale motion and fine-grained textures~\cite{deco, dip, pixeldit}. Inspired by the human cognitive process in biological motion perception~\cite{oliva2006building, troje2008biological, vaina2001functional}, where observers first perceive macroscopic motion and then refine structural and texture details, we design SPEED with a deliberately structured architecture to progressively capture motion, structure, and appearance.
Specifically, we propose a progressive multi-stage Transformer that varies the patch size across layers, decreasing from 64 to 32 and then to 16.
This design enables the model to capture patterns at multiple scales. Shallow layers employ large patches to model global motion, middle layers refine structural alignment, and deeper layers adopt smaller patches to synthesize fine-grained appearance and texture details.
Meanwhile, we introduce a novel Noise-Update-Only (NUO) Attention, which strictly isolates noise updates while providing a fully global spatio-temporal receptive field with a highly efficient complexity, \textit{e.g.,} $\mathcal{O}(3N^2)$ of NUO \textit{v.s} $\mathcal{O}(9N^2)$ of full attention.

\begin{wrapfigure}{r}{0.5\linewidth}
    \centering
    \includegraphics[width=\linewidth]{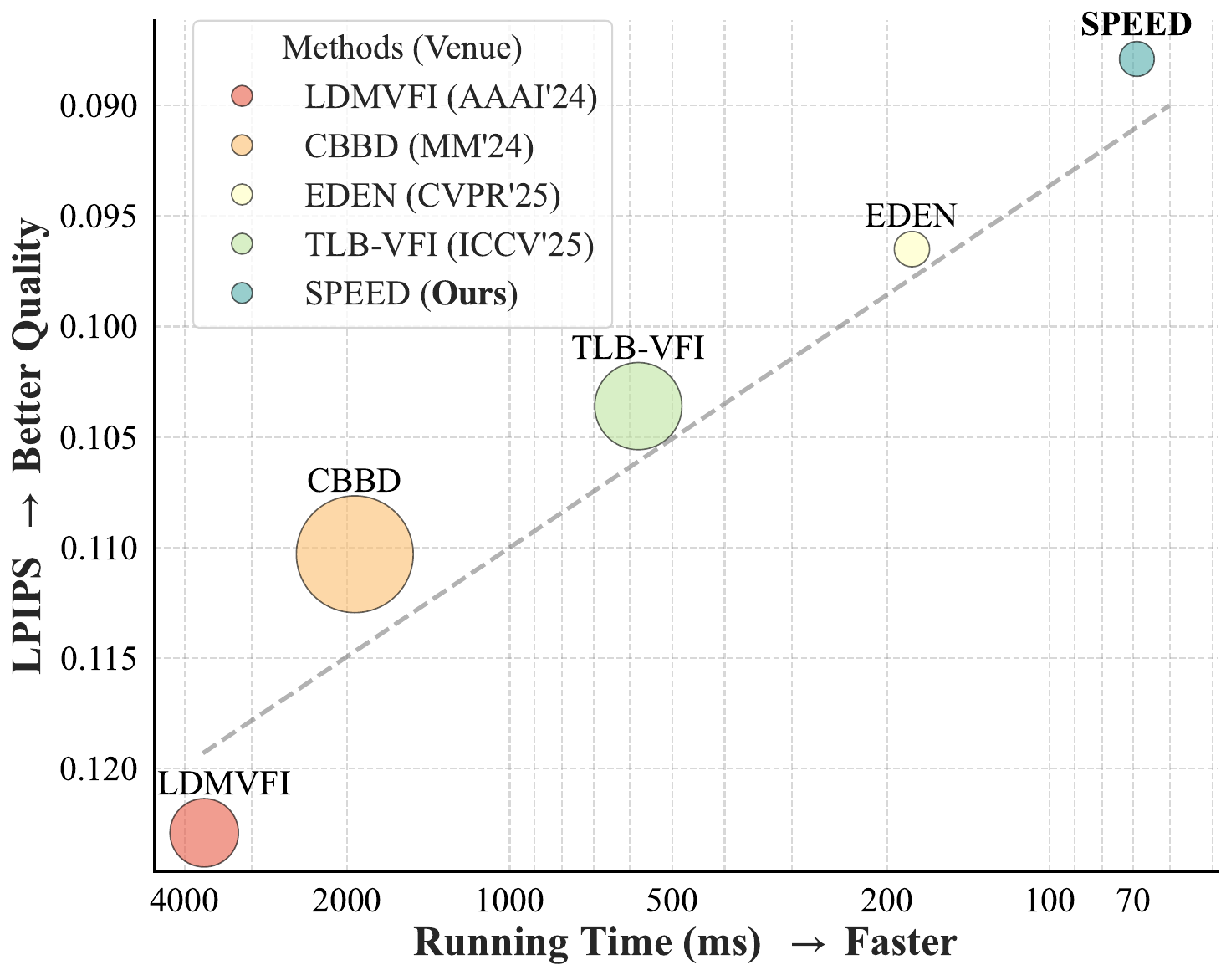}
    \caption{Performance and efficiency comparison of recent diffusion-based VFI methods on SNU-FILM (Extreme)~\cite{snufilm}. The radius indicates the usage of memory.}
    \label{fig:efficiency}
\end{wrapfigure}

Moreover, as one-step generation~\cite{dmd, dmd2, add} has emerged as an effective paradigm for improving inference efficiency, we design Drift-aware Timestep Sampling (DTS) with a dedicated training objective to enable one-step VFI.
By dynamically warping the sampling distribution during training, DTS first learns the complete probability flow ODE trajectory (i.e., multi-step sampling) and then gradually shifts focus toward the boundary point $t=1$. This process acts as a curriculum, allowing the model to smoothly transition from progressive iterative sampling to instantaneous one-step projection.
Besides, we redesign the training objective based on the clean image $x_0$, instead of predicting the noise or velocity field adopted by previous methods~\cite{LDMVFI, LBBDM, VIDIM, eden, tlbvfi}. This formulation enables direct perceptual loss supervision and significantly accelerates convergence.
Extensive experiments on standard and high-resolution benchmarks demonstrate that SPEED significantly outperforms both state-of-the-art optical flow-based models and diffusion baselines in generative quality, inference speed, and memory efficiency. 
For instance, as shown in Fig.~\ref{fig:efficiency}, compared with existing methods, our model achieves the optimal performance (\textbf{0.088} LPIPS) while requiring the least memory consumption (\textbf{10.63\%} lower than EDEN~\cite{eden}) and the lowest time overhead (\textbf{63.32\%} faster than EDEN).

In summary, our main contributions are as follows:
\begin{itemize}
    \item We propose SPEED, a novel one-step diffusion framework that entirely operates in pixel space, explicitly solving the detailed information loss and inference inefficiency bottlenecks of existing latent diffusion VFI models.
    \item We develop a progressive multi-stage architecture that mimics the macroscopic-to-microscopic cognitive process. This dynamic patch size strategy, coupled with an efficient Noise-Update-Only Attention mechanism, ensures stable and robust pixel-space modeling.
    \item We propose Drift-aware Timestep Sampling and a dedicated training objective during training process to facilitate one-step video frame interpolation.
    \item Extensive evaluations on the DAVIS, SNU-FILM, and XTest4K benchmarks demonstrate that the proposed SPEED framework achieves a new state-of-the-art in interpolation quality while reducing inference time and memory usage.
\end{itemize}
\label{sec:intro}

\section{Related Work}

\subsection{Video Frame Interpolation}
Traditional video frame interpolation methods primarily focus on modeling intermediate optical flow, subsequently employing forward~\cite{softmax} or backward warping~\cite{stmfnet, VFIFormer, IFRNet, RIFE, AMT} to synthesize the target frame. For instance, SGM-VFI~\cite{SGMVFI} adopts sparse global matching to refine flow estimation. To specifically address the time-to-location ambiguity caused by non-uniform motions, BiM-VFI~\cite{bimvfi} proposes a bidirectional motion field descriptor paired with a content-aware upsampling network. Recently, state-space models~\cite{mamba} have also gained traction for this task. VFIMamba~\cite{vfimamba} utilizes a global receptive field to capture motion via state-space modeling, while LC-Mamba~\cite{lcmamba} incorporates a shifted local window technique and Hilbert curve-based scanning to precisely capture fine-grained spatiotemporal characteristics. 

Recent works have leveraged the generative capabilities of diffusion models to tackle large-motion VFI. For example, LDMVFI~\cite{LDMVFI} employs a latent diffusion model to generate the intermediate frame conditioned on the given inputs. CBBD~\cite{LBBDM} utilizes a consecutive Brownian bridge diffusion model to synthesize the target frame, while VIDIM~\cite{VIDIM} employs cascaded diffusion models for generation. To better extract temporal information, TLB-VFI~\cite{tlbvfi} introduces a temporal-aware latent Brownian bridge diffusion framework equipped with 3D-wavelet gating. Similarly, EDEN~\cite{eden} replaces the standard VAE with a transformer-based tokenizer, injecting temporal attention and difference embeddings into a diffusion transformer to guide the dynamic motion.

Despite their superior generative priors, these latent diffusion-based methods remain fundamentally bottlenecked by the information loss inherent to VAE compression, as well as the severe inference latency imposed by iterative multi-step sampling. In contrast, our \modelname~discards the VAE entirely, operating within the pixel space. This design explicitly eliminates the information loss, enabling high-quality intermediate frame generation without the constraints of latent compression.

\vspace{-2ex}
\subsection{Diffusion Models}
Denoising Diffusion Probabilistic Models (DDPMs)~\cite{ddpm} have revolutionized generative modeling. LDMs~\cite{ldm} perform the diffusion process within a compressed latent space encoded by a pre-trained VAE~\cite{vae}. This paradigm underpins highly successful frameworks in image and video generation tasks~\cite{magdiff, motionfollower, svd, Latte, sd3, sora, pixart, lstd}. However, despite their success, the latent compression inevitably loses fine-grained information, creating a reconstruction upper bound that degrades fine details and micro-textures—a critical flaw for precise video frame interpolation.

To circumvent the information loss caused by VAEs, recent generalized generative models are actively exploring end-to-end pixel-space diffusion. PixelDiT~\cite{pixeldit}, DiP~\cite{dip}, and DeCo~\cite{deco} demonstrate that Diffusion Transformers~\cite{dit} can effectively model pixel patches without relying on latent compression. Furthermore, JiT~\cite{jit} achieves highly competitive generation with plain Vision Transformers, and PixelNerd~\cite{pixnerd} replaces standard linear projections with Neural Fields to render fine-grained details within large patches. Building upon these cutting-edge advancements, our \modelname~framework tackles large-motion video frame interpolation entirely in the pixel domain. By integrating a progressive multi-stage architecture with an optimized one-step video frame interpolation, we explicitly bypass the VAE bottleneck and achieve high-quality, one-step generation.
\label{sec:related}

\section{Method}

\begin{figure*}[t]
    \centering
    \includegraphics[width=0.95\linewidth]{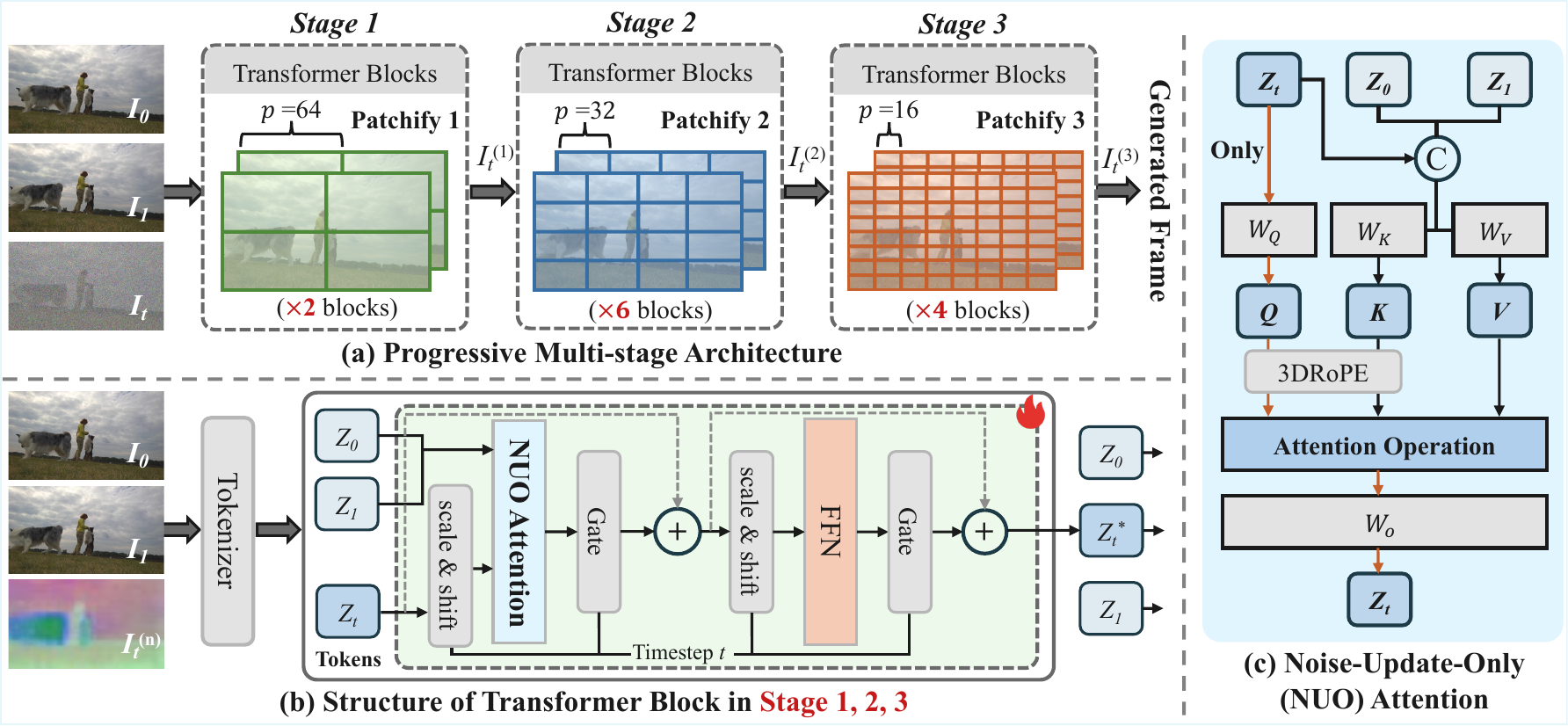}
    \caption{Overview of the proposed \modelname~ framework. The model adopts a three-stage pixel diffusion architecture (a) that progressively transforms motion-aware representations into detail-refined intermediate frames. Each stage consists of several transformer blocks (b) with a deliberately designed Noise-Update-Only (NUO) attention (c) mechanism.}
    \label{fig:pipeline}
\end{figure*}

We propose \modelname, a pixel diffusion framework designed to eliminate the information loss inherent to latent compression and overcome inference inefficiency. As illustrated in Fig.~\ref{fig:pipeline}, our approach operates in the pixel space, relying on three deliberate designs to address the aforementioned challenges: a progressive multi-stage Transformer architecture (Section 3.1) that dynamically scales patch size to sequentially capture motion, structure and appearance; a Noise-Update-Only (NUO) Attention mechanism (Section 3.2) that preserves the deterministic guidance of the condition frames while halving computational overhead; and an optimized one-step generation strategy (Section 3.3) driven by drift-aware timestep sampling and a dedicated training objective.

\subsection{Progressive Multi-stage Architecture}
Given a starting frame $I_0$ and an ending frame $I_1$, \modelname~takes these conditions alongside a noisy intermediate frame $I_t$ and a diffusion timestep $t$ to predict the clean intermediate frame $\bar{I}_t$. Generating frames in pixel space imposes conflicting optimization requirements: the model must simultaneously capture large-scale object displacements and synthesize intricate micro-textures. To resolve this, we propose a deliberately structured architecture designed to progressively capture motion, structure, and appearance. Specifically, we implement a progressive multi-stage Transformer that dynamically varies the patch size across its depth, decreasing from $p$=64 to $p$=32, and finally to $p$=16, as shown in Fig.~\ref{fig:pipeline} (a). This progressive design enables the model to explicitly isolate and capture visual patterns at multiple scales. The shallow layers employ large patches to aggressively discard granular details, forcing the network to focus entirely on modeling global, macroscopic motion trajectories. As the representations propagate to the middle layers, the patch size is reduced to balance the refinement of these motion paths with precise structural alignment. Finally, the deeper layers adopt the smallest patches, allowing the tokens to retain maximal information to synthesize fine-grained appearance and intricate texture details.

The forward computational process within each stage remains architecturally consistent, as shown in Fig.~\ref{fig:pipeline} (b). Before entering the Transformer blocks, the starting frame $I_0$, the ending frame $I_1$, and the noisy intermediate frame $I_t$ are partitioned into non-overlapping patches and linearly projected into sequences of token embeddings via a tokenizer:
\begin{equation}
    \begin{aligned}
    Z_0 &= \text{Tokenizer}(I_0), \\
    Z_1 &= \text{Tokenizer}(I_1), \\
    Z_t &= \text{Tokenizer}(I_t),
    \end{aligned}
\end{equation}
where $Z_{0}, Z_{1}, Z_{t} \in \mathbb{R}^{N \times d}$, with $N$ representing the sequence length ($N = \frac{H \times W}{p^2}$) and $d$ denoting the embedding dimension. A timestep embedding $t_e$ is derived from the current diffusion timestep $t$. Within a generic Transformer block $l$, the forward operation is modulated via adaptive Layer Normalization (adaLN)~\cite{dit} and stabilized by residual connections. 

First, the adaLN module generates the modulation parameters (scale $\alpha$, shift $\beta$, and gate $\gamma$) from $t_e$:
\begin{equation}
    (\alpha_1, \beta_1, \gamma_1, \alpha_2, \beta_2, \gamma_2)= \text{AdaLN}(t_e).
\end{equation}

Next, the embeddings are processed by our specialized NUO attention block (detailed in Section~\ref{sub:nuo}) to produce the updated temporal state $Z_t^{(l)}$. This operation integrates the modulated intermediate frame with the clean boundary conditions:
\begin{equation}
    Z_t^{(l)} = Z_t^{(l)} + \gamma_1 \cdot \text{NUO}(\alpha_1 \cdot \text{LN}(Z_t^{(l)})+\beta_1, Z_0, Z_1).
\end{equation}

Subsequently, this intermediate representation is passed through a Feed-Forward Network (FFN), following the standard Diffusion Transformer block~\cite{dit} formulation, to produce the final output for the subsequent layer $l+1$:
\begin{equation}
    Z_t^{(l+1)} = Z_t^{(l)} + \gamma_2 \cdot \text{FFN}(\alpha_2 \cdot \text{LN}(Z_t^{(l)}) + \beta_2).
\end{equation}

\subsection{Noise-Update-Only (NUO) Attention}
\label{sub:nuo}
Previous methods~\cite{eden, EMA} typically employ a decoupled attention paradigm, combining spatial self-attention for intra-frame modeling with temporal attention for inter-frame interaction. This decoupled architecture inherently restricts the receptive field, as the intermediate frame only attends to contextual information constrained by a localized temporal window. In contrast, our NUO attention mechanism establishes a global receptive field across the entire frame triplet $(I_0, I_t, I_1)$ within a unified attention operation. This ensures that the noisy intermediate frame $I_t$ has unrestricted access to the comprehensive boundary conditions, effectively eliminating inter-frame information bottlenecks.

\noindent \textbf{NUO Formulation.} In standard spatio-temporal full attention, information flows bidirectionally among all input tokens. This bidirectional exchange causally corrupts the condition representations with noise, severely degrading the deterministic guidance required for high-quality interpolation. To explicitly prevent this corruption, we propose the NUO Attention. The overall computational pipeline is illustrated in Fig.~\ref{fig:pipeline} (c).

Specifically, we generate distinct queries, keys, and values from the input hidden states $Z_0, Z_t^{(l)}, \text{ and } Z_1$. Crucially, the Query $Q_t$ is extracted solely from the noisy target $Z_t^{(l)}$, whereas the Keys $K_{0,1,t}$ and Values $V_{0,1,t}$ aggregate the global context across all three frames:
\begin{equation}
    \begin{gathered}
    Q_t = Z_t^{(l)} \cdot W_Q, \\
    K_{0,1,t} = \text{Concat}(Z_0, Z_t^{(l)}, Z_1) \cdot W_K, \\
    V_{0,1,t} = \text{Concat}(Z_0, Z_t^{(l)}, Z_1) \cdot W_V.
    \end{gathered}
\end{equation}

To inject precise spatio-temporal awareness into these tokens, we apply 3D Rotary Position Embedding (RoPE)~\cite{wan} to the projected vectors. This explicitly encodes their spatial coordinates $(h,w)$ and temporal origin $t' \in \{0, t, 1\}$ without altering the underlying feature magnitudes:
\begin{equation}
    \begin{gathered}
    \tilde{Q}_t = R_{\theta, p_t} \cdot Q_t, \\
    \tilde{K}_{0,1,t} = \text{Concat}(R_{\theta, p_0}, R_{\theta, p_t}, R_{\theta, p_1}) \cdot K_{0,1,t}.
    \end{gathered}
\end{equation}

Finally, the cross-frame attention output for block $l$ is computed. The isolated query ensures that the resulting attention matrix strictly maps the global context back to $Z_t^{(l)}$:
\begin{equation}
    \begin{gathered}
    Z_t^{(l)} = \text{Softmax}\left(\frac{\tilde{Q}_t \cdot \tilde{K}_{0,1,t}^T}{\sqrt{d}}\right) \cdot V_{0,1,t}, \\
    Z_t^{(l)} = Z_t^{(l)} \cdot W_O.
    \end{gathered}
\end{equation}

Due to this strictly asymmetric query derivation, NUO exclusively updates $Z_t^{(l)}$ while referencing the complete, uncorrupted spatio-temporal context of the condition frames. Furthermore, by calculating attention for $N$ queries against $3N$ keys—rather than the standard $3N$ queries against $3N$ keys—this design dramatically reduces the computational complexity of the attention operation from $\mathcal{O}(9N^2)$ down to $\mathcal{O}(3N^2)$. This algorithmic efficiency facilitates the remarkable inference speed of our framework.

\subsection{One-Step Video Frame Interpolation}

\begin{wrapfigure}{r}{0.5\linewidth}
    \centering
    \includegraphics[width=1.0\linewidth]{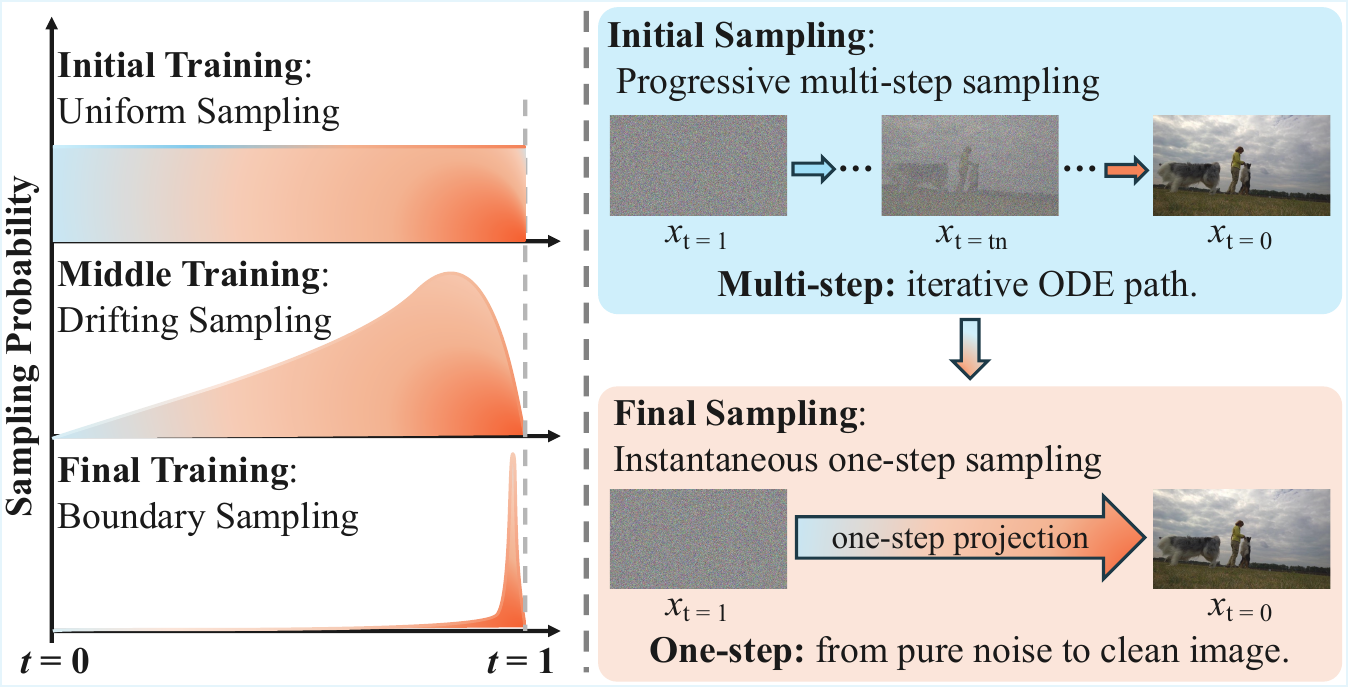}
    \caption{Illustration of Drift-aware Timestep Sampling.}
    \label{fig:timestep_density_drift}
\end{wrapfigure}

\noindent \textbf{Drift-aware Timestep Sampling.} To achieve high-quality one-step generation without the performance decay typical of rapid diffusion inference, we introduce a drift-aware timestep sampling, as show in Fig.~\ref{fig:timestep_density_drift}.
Let $p \in [0, 1)$ denote the training progress, and $t \sim \mathcal{U}(0, 1)$ be the initial uniform time step. We define a drift coefficient $s(p)$ and map $t$ to a warped timestep $t'$:
\begin{gather}
    s(p) = \frac{1}{1-p}, \\
    t' = \frac{s(p) \cdot t}{1 + (s(p) - 1) \cdot t}
\end{gather}

The initial uniform sampling ($p=0, s(0)=1 \rightarrow t'=t$) allows the model to explore the full probability flow ODE trajectory, establishing a robust multi-step understanding of the data manifold. As training progresses ($p \to 1$), we shift optimization focus to the most challenging condition for one-step inference: $t=1$. As $s(p) \to \infty$:
\begin{equation}
    \lim_{s \to \infty} t' = \lim_{s \to \infty} \frac{s \cdot t}{s \cdot t + 1 - t} = 1 \quad (\forall t > 0).
\end{equation}

This forces the computational power and optimization priority to collapse toward the $t=1$ region. The curriculum perfectly transitions the network from learning progressive sampling to mastering instantaneous one-step projection from noise.

\noindent \textbf{One-step VFI Objectives.} For VFI, the target frame $I_t$ (our clean target $x_0$) shares dense structural consistency with the input conditions $I_0$ and $I_1$. Directly predicting $x_0$ allows the network to heavily leverage this powerful structural prior. In contrast, previous methods~\cite{LDMVFI, VIDIM, LBBDM, hifi, eden, tlbvfi} utilize standard formulations, which predict an erratic noise $\epsilon$ or velocity field $v$ (e.g., $v = x_0 - \epsilon$) discard this correlation. Forcing the network to map highly structured conditions to a chaotic noise vector is computationally demanding and prone to severe instability.

\begin{wrapfigure}{r}{0.5\linewidth}
\centering
\includegraphics[width=\linewidth]{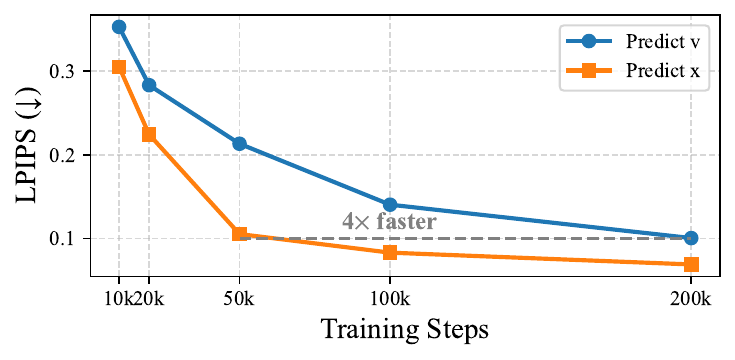} 
\caption{Convergence comparison between predicting the velocity field ($v$) and predicting the clean target ($x$). Evaluated on DAVIS-256~\cite{davis} across 200k training steps.}
\label{fig:pred_target}
\end{wrapfigure}

While the $v$-prediction model struggles to map erratic, high-dimensional velocity vectors, the $x_0$-prediction model rapidly establishes structural awareness. As illustrated in Fig.~\ref{fig:pred_target}, by exploiting the VFI-specific structural prior and shifting the prediction objective to directly recover the clean target $x_0$, we significantly reduce optimization difficulty, yielding a substantially faster convergence rate and superior final generative quality.

Furthermore, our direct modeling in pixel space provides a critical advantage over latent diffusion models: we completely avoid the need to pass representations through a VAE decoder to compute image-level metrics during the diffusion training loop. This allows us to directly apply perceptual and structural constraints to our training target to further enhance visual quality. 

The total prediction objective is formulated as a weighted combination of the pixel-wise reconstruction loss ($\mathcal{L}_{x_0}$), perceptual loss ($\mathcal{L}_{\text{LPIPS}}$)~\cite{lpips}, and style loss ($\mathcal{L}_{\text{style}}$)~\cite{FILM}:
\begin{equation}
\label{eq:loss_func}
\begin{gathered}
    \mathcal{L}_{x_0} = \mathbb{E}_{x_0, I_0, I_1, \epsilon, t} \left\| x_0 - f_\theta(I_t, I_0, I_1, t) \right\|_2^2, \\
    \mathcal{L}_{\text{total}} = w_1 \mathcal{L}_{x_0} + w_l \mathcal{L}_{\text{LPIPS}} + w_s \mathcal{L}_{\text{style}}
\end{gathered}
\end{equation}

\noindent \textbf{One-step VFI Inference.} DTS and one-step VFI objectives endow the model with exceptional robustness and adaptability to high-noise regimes ($t=1$). During inference, we completely bypass traditional iterative sampling. Instead, we sample a standard Gaussian noise vector $\epsilon \sim \mathcal{N}(0, \mathbf{I})$ and set the timestep to $t=1$. The intermediate frame $\tilde{I}_t$ is then generated instantaneously, conditioned on frames $I_0$ and $I_1$:
\begin{equation}
    \tilde{I}_t = f_\theta(\epsilon, I_0, I_1, t=1).
\end{equation}

This straightforward formulation leverages the optimized boundary projection capabilities of \modelname, enabling instantaneous, high-quality frame interpolation in one step. We provide more details in the supplementary materials.
\label{sec:methods}

\section{Experiments}
\begin{table*}[t]
\centering
\caption{Quantitative evaluation across different challenging benchmarks, including DAVIS~\cite{davis} and the Extreme and Hard subsets of SNU-FILM~\cite{snufilm}. The best results are highlighted in \textbf{bold}, and the second-best are \underline{underlined}.}
\resizebox{\textwidth}{!}{
\begin{tabular}{lcccccccccc}
\toprule
\multirow{3}{*}{Method} & \multicolumn{4}{c}{\textbf{DAVIS}} & \multicolumn{6}{c}{\textbf{SNU-FILM}} \\
\cmidrule(lr){2-5} \cmidrule(lr){6-11}
& \multirow{2}{*}{LPIPS$\downarrow$} & \multirow{2}{*}{FloLPIPS$\downarrow$} & \multirow{2}{*}{RT (ms)$\downarrow$} & \multirow{2}{*}{Mem (G)$\downarrow$} & \multicolumn{2}{c}{Hard} & \multicolumn{2}{c}{Extreme} & \multirow{2}{*}{RT (ms)$\downarrow$} & \multirow{2}{*}{Mem (G)$\downarrow$} \\
\cmidrule(lr){6-7} \cmidrule(lr){8-9}
& & & & & LPIPS$\downarrow$ & FloLPIPS$\downarrow$ & LPIPS$\downarrow$ & FloLPIPS$\downarrow$ & & \\
\midrule
% XVFI'21~\cite{XVFI}         & 0.1253 & 0.1761 & 83.5 & \underline{2.526} & 0.0686 & 0.1174 & 0.1203 & 0.1927 & 141.2 & 4.100 \\
SGM-VFI~\cite{SGMVFI}CVPR'24    & 0.1121 & 0.1528 & 106.3 & 3.100 & 0.0610 & 0.0995 & 0.1182 & 0.1876 & 197.4 & 3.970 \\
VFIMamba~\cite{vfimamba}NIPS'24 & 0.1092 & 0.1484 & \underline{53.2} & 3.456 & 0.0615 & 0.0993 & 0.1160 & 0.1861 & \underline{115.0} & 5.122 \\
LC-Mamba~\cite{lcmamba}CVPR'25  & 0.1278 & 0.1741 & 78.9 & 4.428 & 0.0709 & 0.1160 & 0.1350 & 0.2148 & 131.1 & 7.872 \\
BiM-VFI~\cite{bimvfi}CVPR'25    & 0.0894 & 0.1270 & 139.9 & \underline{2.594} & 0.0535 & 0.0854 & 0.0981 & \underline{0.1556} & 213.1 & 4.412 \\
LDMVFI~\cite{LDMVFI}AAAI'24     & 0.1101 & 0.1543 & 3551.9 & 5.018 & 0.0602 & 0.1141 & 0.1229 & 0.2041 & 3678.7 & 7.346 \\
CBBD~\cite{LBBDM}MM'24        & 0.0968 & 0.1358 & 1643.6 & 7.014 & 0.0489 & 0.0909 & 0.1103 & 0.1839 & 1935.5 & 19.060 \\
EDEN~\cite{eden}CVPR'25         & \underline{0.0856} & \underline{0.1179} & 87.7 & 2.666 & \underline{0.0483} & 0.0882 & \underline{0.0965} & 0.1623 & 189.5 & \underline{2.860} \\
TLB-VFI~\cite{tlbvfi}ICCV'25    & 0.0914 & 0.1260 & 293.1 & 4.218 & 0.0501 & \underline{0.0847} & 0.1036 & 0.1619 & 577.8 & 11.164 \\
\midrule
\textbf{Ours}               & \textbf{0.0831} & \textbf{0.1155} & \textbf{36.6} & \textbf{2.268} & \textbf{0.0476} & \textbf{0.0802} & \textbf{0.0880} & \textbf{0.1447} & \textbf{69.5} & \textbf{2.556} \\
\bottomrule
\end{tabular}
}
\label{tab:comparisons}
\end{table*}

\subsection{Datasets and Evaluation Metrics}
Following recent diffusion-based video frame interpolation works~\cite{VIDIM, hifi, eden, LDMVFI, LBBDM, tlbvfi}, we train our model on the large-scale LAVIB dataset~\cite{lavib} and evaluate its performance across three representative benchmarks: DAVIS~\cite{davis} and SNU-FILM~\cite{snufilm} for complex non-linear motions, and XTest~\cite{XVFI} for high-resolution scenarios. To accurately reflect visual quality, we evaluate generation quality using perceptual metrics including LPIPS~\cite{lpips} and FloLPIPS~\cite{flolpips}. Furthermore, to rigorously benchmark the practical advantages of our one-step VFI, we report the Running Time (RT) and Peak Memory (Mem) required to generate a single intermediate frame. Efficiency metrics (RT and Mem) are evaluated on an NVIDIA A100-80G GPU at resolutions of 480$\times$854 (DAVIS), 720$\times$1280 (SNU-FILM), and 2160$\times$4096 (XTest4K). Comprehensive implementation details, including network configurations and hyperparameter settings, are provided in the supplementary materials.

\subsection{Comparisons with Previous Works}
To comprehensively evaluate \modelname, we compare it against optical flow-based methods (XVFI~\cite{XVFI}, SGM-VFI~\cite{SGMVFI}, VFIMamba~\cite{vfimamba}, LC-Mamba~\cite{lcmamba}, BiM-VFI~\cite{bimvfi}) and state-of-the-art diffusion-based models (LDMVFI~\cite{LDMVFI}, CBBD~\cite{LBBDM}, EDEN~\cite{eden}, TLB-VFI~\cite{tlbvfi}). 

\noindent \textbf{Performance on Large Motion Benchmarks.} As shown in Table~\ref{tab:comparisons}, \modelname~demonstrates exceptional robustness on the highly dynamic DAVIS and SNU-FILM datasets, sweeping all evaluated metrics. In terms of generative quality, our method achieves an LPIPS of 0.0880 and a FloLPIPS of 0.1447 on the challenging SNU-FILM Extreme subset, yielding an 8.8\% and 10.8\% improvement over the previous best diffusion model, EDEN, respectively. It also consistently outperforms BiM-VFI, establishing a new absolute state-of-the-art in perceptual quality. 

Crucially, \modelname~fundamentally resolves the prohibitive latency of diffusion models while simultaneously dominating highly optimized optical flow-based methods. On DAVIS, it generates a frame in just 36.6 ms. This represents a staggering 58.3\% reduction in inference time compared to EDEN, and it is 73.8\% and 31.2\% faster than BiM-VFI and VFIMamba, respectively. Alongside these speedups, \modelname~maintains the lowest peak memory usage (2.268G) among all evaluated methods.

\begin{wraptable}{r}{0.5\linewidth}
\centering
\caption{Quantitative evaluation on XTest4K~\cite{XVFI}. The best results are in \textbf{bold}, and the second best are \underline{underlined}. TLB-VFI~\cite{tlbvfi} is the latest VFI model.}
\resizebox{\linewidth}{!}{
\begin{tabular}{lcccc}
\toprule
& \multicolumn{4}{c}{\textbf{XTest4K}} \\
\cmidrule(lr){2-5}
Method 
& LPIPS$\downarrow$ 
& FloLPIPS$\downarrow$ 
& RT (ms)$\downarrow$ 
& Mem (G)$\downarrow$ \\
\midrule
XVFI~\cite{XVFI}ICCV'21        & \underline{0.1099} & 0.1982 & 1137.1 & 18.036 \\
SGM-VFI~\cite{SGMVFI}CVPR'24  & 0.1199 & \underline{0.1784} & \textbf{758.1} & 13.858 \\
% VFIMamba~\cite{vfimamba} & 0.1272 & 0.1935 & 1.044 & 31.394 \\
% LC-Mamba~\cite{lcmamba}  & 0.1696 & 0.2300 & 1.187 & 56.448 \\
% BiM-VFI~\cite{bimvfi}    & \textbf{0.0905} & \textbf{0.1269} & 1.658 & 39.262 \\
LDMVFI~\cite{LDMVFI}AAAI'24     & 0.2469 & 0.3331 & 7292.7 & 27.344 \\
CBBD~\cite{LBBDM}MM'24        & OOM & OOM & OOM & OOM \\
EDEN~\cite{eden}CVPR'25         & 0.2426 & 0.3391 & 1426.2 & \underline{7.418} \\
TLB-VFI~\cite{tlbvfi}ICCV'25    & 0.2207 & 0.3162 & 3273.6 & 72.414 \\
\midrule
\textbf{Ours} 
& \textbf{0.1070} 
& \textbf{0.1484} 
& \underline{959.4} 
& \textbf{7.210} \\
\bottomrule
\end{tabular}
}
\label{tab:xtest4k}
\end{wraptable}

\noindent \textbf{Performance on High-Resolution Scenarios (4K).} Interpolating at high resolutions reveals the architectural bottlenecks inherent in latent compression. As presented in Table~\ref{tab:xtest4k}, existing diffusion models struggle severely at $2160 \times 4096$ resolution: CBBD fails due to Out-Of-Memory (OOM) errors, while LDMVFI, TLB-VFI, and EDEN suffer severe perceptual degradation (all yielding LPIPS $> 0.22$). By operating in pixel space, \modelname~avoids information loss entirely. It not only achieves the highest perceptual quality (LPIPS of 0.1070) and structural accuracy (FloLPIPS of 0.1484)—surpassing even 4K-specific specialists like XVFI and SGM-VFI—but it also does so with the lowest memory usage (7.210G), using nearly half the memory required by SGM-VFI.

\subsection{Visualization}

\begin{figure*}[t]
\centering
\includegraphics[width=0.95\textwidth]{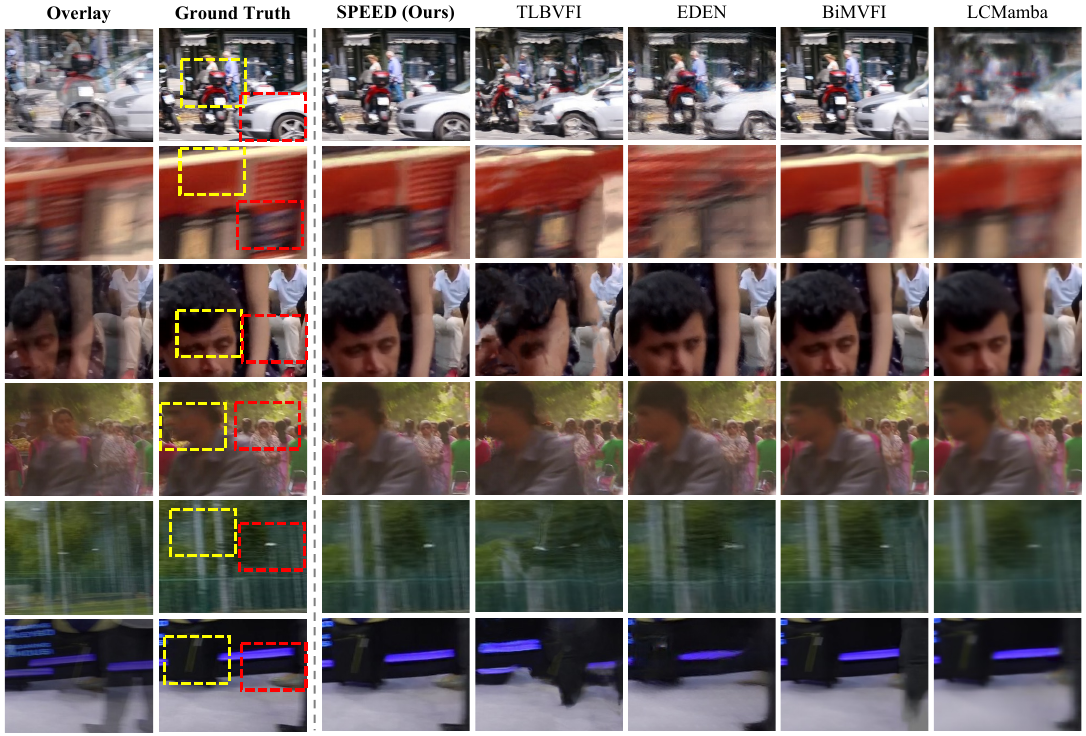}
\caption{Visual comparison of SPEED against state-of-the-art methods on challenging large-motion and complex textures scenarios. The yellow and red dashed boxes highlight the regions where we perform significantly better. SPEED consistently outperforms both optical flow-based models and latent diffusion baselines in preserving structural integrity and sharp micro-textures.\looseness=-1}
\label{fig:visualization}
\end{figure*}

Fig.~\ref{fig:visualization} presents a qualitative visual comparison between our proposed SPEED framework and recent state-of-the-art methods, including both advanced optical flow-based models (LC-Mamba, BiM-VFI) and latent diffusion models (EDEN, TLB-VFI). 

As illustrated in the third and fifth examples—which feature intricate high-frequency details such as human facial features and the structural grid lines of a fence—previous latent diffusion methods (EDEN and TLB-VFI) suffer from severe blurring and structural distortion. This visually corroborates our assertion regarding the information loss inherent to VAE compression. In contrast, by operating entirely in the pixel space and utilizing a fine-grained stage ($p$=16), SPEED effectively synthesizes these high-frequency micro-textures, generating notably sharper and more realistic details.

Furthermore, in scenarios involving highly dynamic and large, non-linear motions—such as the rapidly moving vehicle in the first example and the pedestrian in the fourth example—previous flow-based and diffusion methods struggle to accurately align structural boundaries, often resulting in severe ghosting artifacts, tearing, or smeared textures. This limitation is particularly pronounced in the second and sixth examples, which feature complex non-linear motion occurring precisely at the frame boundaries. When a portion of this boundary motion is missing or abruptly truncated between the input frames, existing models fail to infer the correct trajectory. Our method, however, successfully tracks the macroscopic motion and synthesizes the missing dynamics, generating highly reasonable and structurally coherent intermediate frames that most closely align with the Ground Truth (GT). We provide more visualization results in the supplementary materials.

\subsection{Ablation Study}
To validate the core components of \modelname, we train all variants from scratch for 200k steps, tailoring evaluation resolutions to specific ablation objectives. For \textbf{structural designs} (the progressive architecture and NUO attention), we evaluate on the 720p SNU-FILM (Extreme) dataset. This high resolution fully accommodates our large patch sizes and prevents architectural improvements from being obscured by limited spatial dimensions. Conversely, for resolution-independent \textbf{training strategies} (e.g., one-step VFI), we evaluate on the downsampled DAVIS-256 dataset (144$\times$256) to maximize computational efficiency while maintaining reliable validation.

\begin{wraptable}{r}{0.5\linewidth}
\centering
\caption{Ablation study on multi-stage architecture configurations, evaluated on the SNU-FILM (Extreme)~\cite{snufilm} subset. The sequence \textit{X-Y-Z} denotes the number of transformer blocks at patch sizes $p$=64, $p$=32, and $p$=16, respectively.}
\resizebox{\linewidth}{!}{
\begin{tabular}{llccc}
\toprule
\multirow{2}{*}{\textbf{Setting}} & \multirow{2}{*}{\textbf{Configuration}} & \multicolumn{3}{c}{\textbf{SNU-FILM (Extreme)}} \\
\cmidrule(lr){3-5}
 &  & LPIPS$\downarrow$ & FloLPIPS$\downarrow$ & RT (ms)$\downarrow$ \\
\midrule
\multirow{2}{*}{w/o multi-stage} 
 & $p$=32 (12 layers) & 0.1071 & 0.1876 & \textbf{68.2} \\
 & $p$=16 (12 layers) & 0.0960 & 0.1611 & 142.3 \\
\midrule
\multirow{4}{*}{w/ multi-stage} 
 & 4-6-2  & 0.0983 & 0.1624 & \underline{69.2} \\
 & 3-6-3  & 0.0964 & 0.1619 & 69.5 \\
 & 2-4-6  & \textbf{0.0945} & \textbf{0.1553} & 89.7 \\
 & \textbf{2-6-4 (Ours)} & \underline{0.0956} & \underline{0.1581} & 69.5 \\
\bottomrule
\end{tabular}}
\label{tab:ablation_multistage}
\end{wraptable}

\noindent \textbf{Effectiveness of the Progressive Multi-Stage Architecture.}
We evaluate our progressive multi-stage architecture against single-stage baselines, as show in Table~\ref{tab:ablation_multistage}. A strictly coarse patch size ($p$=32) accelerates inference but degrades perceptual quality due to lacking the token granularity necessary for high-frequency textures. Conversely, a strictly fine patch size ($p$=16) improves quality but imposes a severe computational bottleneck.

By introducing the progressive multi-stage architecture, we enforce a macroscopic-to-microscopic modeling curriculum. Among the tested configurations, placing more weight on early stages ``4-6-2'' lacks sufficient depth for micro-textures, while prioritizing fine-grained stages ``2-4-6'' yields high quality but slows down inference. Therefore, we adopt the ``2-6-4'' configuration as the optimal trade-off, achieving comparable perceptual quality to the "$p$=16" baseline at less than half the computational cost.

\noindent \textbf{Effectiveness of Noise-Update-Only Attention.}
We compare our NUO attention against two baselines: \textit{(Self-Attn + Temp-Attn)} and full-attention. As shown in Table~\ref{tab:nuo_ablation}, \textit{Self-Attn + Temp-Attn} yields the worst perceptual quality, cause the limited spatio-temporal receptive field. The full-attention baseline achieves a global receptive field, improving motion alignment. However, it incurs a massive computational cost
\begin{wraptable}{r}{0.5\linewidth}
\centering
\caption{Ablation study of the attention mechanism evaluated on the SNU-FILM (Extreme) subset~\cite{snufilm}.}
\resizebox{\linewidth}{!}{
\begin{tabular}{l cccc}
\toprule
\multirow{2}{*}{\textbf{Attention Design}} & \multicolumn{4}{c}{\textbf{SNU-FILM (Extreme)}} \\
\cmidrule(lr){2-5}
 & LPIPS$\downarrow$ & FloLPIPS$\downarrow$ & RT (ms)$\downarrow$ & Mem (G)$\downarrow$ \\
\midrule
Self-Attn + Temp-Attn      & 0.0997 & 0.1683 & 78.6  & 2.736 \\ 
Full-Attn & 0.0960 & 0.1596 & 130.7 & 2.620 \\
\midrule
\textbf{NUO-Attn (Ours)}   & \textbf{0.0956} & \textbf{0.1581} & \textbf{69.5}  & \textbf{2.556} \\
\bottomrule
\end{tabular}}
\label{tab:nuo_ablation}
\end{wraptable}
and suffers a performance penalty because noise from the intermediate frame $I_t$ inevitably leaks into the clean condition frames $I_0$ and $I_1$ during the attention operation.

By confining queries strictly to $I_t$, our NUO attention effectively filters out this noise leakage while maintaining the global receptive field necessary for capturing large motions. Furthermore, it nearly halves the inference time compared to full attention and establishes the lowest memory usage, proving it is both a structural necessity and a critical efficiency optimization.

\begin{wraptable}{r}{0.5\linewidth}
\centering
\caption{Ablation study of one-step VFI evaluated on DAVIS-256~\cite{davis}.}
\resizebox{\linewidth}{!}{
\begin{tabular}{cccccc}
\toprule
\multirow{2}{*}{\textbf{Model}} & \multirow{2}{*}{\textbf{predict\_v}} & \multirow{2}{*}{\textbf{predict\_x}} & \multirow{2}{*}{\textbf{DTS}} & \multicolumn{2}{c}{\textbf{DAVIS-256}} \\
\cmidrule(lr){5-6}
 & & & & LPIPS$\downarrow$ & FloLPIPS$\downarrow$ \\ 
\midrule
\multirow{4}{*}{SPEED} & \Checkmark &   \XSolidBrush      & \XSolidBrush & 0.1184             & 0.1529             \\
                       & \Checkmark &   \XSolidBrush      & \Checkmark  & 0.1055             & 0.1372             \\
\cmidrule(lr){2-6}
                       &    \XSolidBrush     & \Checkmark & \XSolidBrush & \underline{0.0885} & \underline{0.1268} \\
                       &   \XSolidBrush      & \Checkmark & \Checkmark  & \textbf{0.0639}    & \textbf{0.0931}    \\ 
\bottomrule
\end{tabular}}
\label{tab:ab_predict_x_v}
\end{wraptable}

\noindent \textbf{Effectiveness of One-step Inference.} We ablate the interaction between our proposed Drift-aware Timestep Sampling (DTS) and the prediction target ($v$ vs. $x_0$). As shown in Table~\ref{tab:ab_predict_x_v}, predicting $v$ yields poor perceptual quality (LPIPS of 0.1184). While adding DTS improves this, the erratic nature of $v$ remains a fundamental bottleneck. Conversely, shifting the target to $x_0$ significantly reduces optimization difficulty, improving LPIPS to 0.0885 even without DTS. However, the most profound improvement occurs when combining the $x_0$ prediction with DTS. This synergistic combination achieves an outstanding LPIPS of 0.0639 and FloLPIPS of 0.0931, confirming that explicit $x_0$ supervision and DTS are both critical for effective one-step generation. We also provide experimental results on latent diffusion in the supplementary materials.

\begin{figure}[h]
\centering
\includegraphics[width=0.9\linewidth]{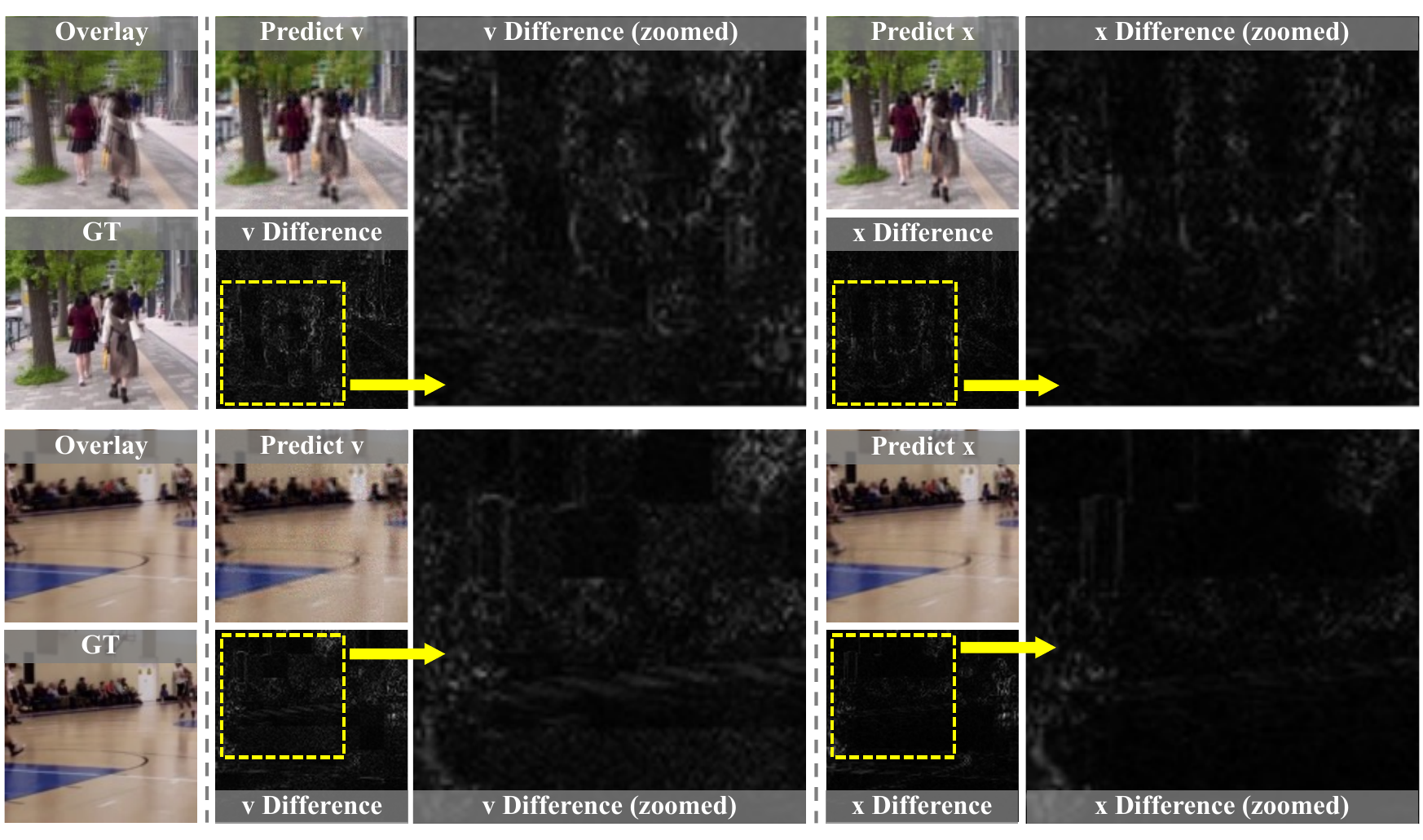}
    \caption{Visual comparison of different prediction targets. The latter four columns visualize the differences between the generated intermediate frames and the ground truth, alongside their magnified local details.}
    \label{fig:predict_v_x}
\end{figure}

\noindent \textbf{Qualitative Effect of SPEED Objectives.} As illustrated in Fig.~\ref{fig:predict_v_x}, models predicting $v$ struggle to effectively leverage the condition frames, resulting in severe noise and local structural degradation. By exploiting the VFI-specific structural prior and shifting the objective to recover the $x_0$, we significantly reduce optimization difficulty, yielding markedly cleaner and sharper predictions.
\label{sec:exps}

\section{Conclusion}
In this paper, we proposed \modelname, a one-step pixel diffusion framework that fundamentally addresses the fine-grained detail loss and inference inefficiency inherent to existing diffusion models for VFI. We introduced a progressive multi-stage architecture with dynamic patch scaling to seamlessly capture multi-scale motion, structure, and appearance representations. Furthermore, we designed a novel Noise-Update-Only (NUO) Attention mechanism, which prevents the semantic degradation of clean condition frames while reducing computational overhead by nearly 50\%. To unlock high-fidelity one-step inference, we proposed Drift-aware Timestep Sampling (DTS) strategy and a tailored training objective. Extensive evaluations demonstrate that \modelname~establishes a new state-of-the-art across standard and extreme 4K benchmarks, delivering superior perceptual quality while drastically reducing inference latency and memory consumption.
\label{sec:conclusion}

\clearpage

\bibliographystyle{plainnat}
\bibliography{main}

@inproceedings{RIFE,
  title={Real-time intermediate flow estimation for video frame interpolation},
  author={Huang, Zhewei and Zhang, Tianyuan and Heng, Wen and Shi, Boxin and Zhou, Shuchang},
  booktitle={European Conference on Computer Vision},
  pages={624--642},
  year={2022},
}

@inproceedings{EMA,
  title={Extracting motion and appearance via inter-frame attention for efficient video frame interpolation},
  author={Zhang, Guozhen and Zhu, Yuhan and Wang, Haonan and Chen, Youxin and Wu, Gangshan and Wang, Limin},
  booktitle={Proceedings of the IEEE/CVF Conference on Computer Vision and Pattern Recognition},
  pages={5682--5692},
  year={2023}
}

@inproceedings{XVFI,
  title={Xvfi: extreme video frame interpolation},
  author={Sim, Hyeonjun and Oh, Jihyong and Kim, Munchurl},
  booktitle={Proceedings of the IEEE/CVF international conference on computer vision},
  pages={14489--14498},
  year={2021}
}

@inproceedings{FILM,
  title={Film: Frame interpolation for large motion},
  author={Reda, Fitsum and Kontkanen, Janne and Tabellion, Eric and Sun, Deqing and Pantofaru, Caroline and Curless, Brian},
  booktitle={European Conference on Computer Vision},
  pages={250--266},
  year={2022}
}

@inproceedings{LearningTo,
  title={Learning to synthesize motion blur},
  author={Brooks, Tim and Barron, Jonathan T},
  booktitle={Proceedings of the IEEE/CVF Conference on Computer Vision and Pattern Recognition},
  pages={6840--6848},
  year={2019}
}

@inproceedings{NCM,
  title={Neighbor correspondence matching for flow-based video frame synthesis},
  author={Jia, Zhaoyang and Lu, Yan and Li, Houqiang},
  booktitle={Proceedings of the 30th ACM International Conference on Multimedia},
  pages={5389--5397},
  year={2022}
}

@inproceedings{IFRNet,
  title={Ifrnet: Intermediate feature refine network for efficient frame interpolation},
  author={Kong, Lingtong and Jiang, Boyuan and Luo, Donghao and Chu, Wenqing and Huang, Xiaoming and Tai, Ying and Wang, Chengjie and Yang, Jie},
  booktitle={Proceedings of the IEEE/CVF Conference on Computer Vision and Pattern Recognition},
  pages={1969--1978},
  year={2022}
}

@inproceedings{VFIFormer,
  title={Video frame interpolation with transformer},
  author={Lu, Liying and Wu, Ruizheng and Lin, Huaijia and Lu, Jiangbo and Jia, Jiaya},
  booktitle={Proceedings of the IEEE/CVF Conference on Computer Vision and Pattern Recognition},
  pages={3532--3542},
  year={2022}
}

@inproceedings{slomo,
  title={Super slomo: High quality estimation of multiple intermediate frames for video interpolation},
  author={Jiang, Huaizu and Sun, Deqing and Jampani, Varun and Yang, Ming-Hsuan and Learned-Miller, Erik and Kautz, Jan},
  booktitle={Proceedings of the IEEE conference on computer vision and pattern recognition},
  pages={9000--9008},
  year={2018}
}

@inproceedings{softmax,
  title={Softmax splatting for video frame interpolation},
  author={Niklaus, Simon and Liu, Feng},
  booktitle={Proceedings of the IEEE/CVF Conference on Computer Vision and Pattern Recognition},
  pages={5437--5446},
  year={2020}
}

@inproceedings{deepstereo,
  title={Deepstereo: Learning to predict new views from the world's imagery},
  author={Flynn, John and Neulander, Ivan and Philbin, James and Snavely, Noah},
  booktitle={Proceedings of the IEEE conference on computer vision and pattern recognition},
  pages={5515--5524},
  year={2016}
}

@inproceedings{AMT,
  title={AMT: All-Pairs Multi-Field Transforms for Efficient Frame Interpolation},
  author={Li, Zhen and Zhu, Zuo-Liang and Han, Ling-Hao and Hou, Qibin and Guo, Chun-Le and Cheng, Ming-Ming},
  booktitle={Proceedings of the IEEE/CVF Conference on Computer Vision and Pattern Recognition},
  pages={9801--9810},
  year={2023}
}

@inproceedings{vidcom,
  title={Video compression through image interpolation},
  author={Wu, Chao-Yuan and Singhal, Nayan and Krahenbuhl, Philipp},
  booktitle={Proceedings of the European conference on computer vision (ECCV)},
  pages={416--431},
  year={2018}
}

@inproceedings{safa,
  title={Scale-Adaptive Feature Aggregation for Efficient Space-Time Video Super-Resolution},
  author={Huang, Zhewei and Huang, Ailin and Hu, Xiaotao and Hu, Chen and Xu, Jun and Zhou, Shuchang},
  booktitle={Winter Conference on Applications of Computer Vision},
  year={2024}
}

@inproceedings{blur,
  title={Blur Interpolation Transformer for Real-World Motion from Blur},
  author={Zhong, Zhihang and Cao, Mingdeng and Ji, Xiang and Zheng, Yinqiang and Sato, Imari},
  booktitle={Proceedings of the IEEE/CVF Conference on Computer Vision and Pattern Recognition},
  pages={5713--5723},
  year={2023}
}

@inproceedings{videoinr,
  title={Videoinr: Learning video implicit neural representation for continuous space-time super-resolution},
  author={Chen, Zeyuan and Chen, Yinbo and Liu, Jingwen and Xu, Xingqian and Goel, Vidit and Wang, Zhangyang and Shi, Humphrey and Wang, Xiaolong},
  booktitle={Proceedings of the IEEE/CVF Conference on Computer Vision and Pattern Recognition},
  pages={2047--2057},
  year={2022}
}

@inproceedings{SGMVFI,
  title={Sparse Global Matching for Video Frame Interpolation with Large Motion},
  author={Liu, Chunxu and Zhang, Guozhen and Zhao, Rui and Wang, Limin},
  booktitle={Proceedings of the IEEE/CVF Conference on Computer Vision and Pattern Recognition},
  pages={19125--19134},
  year={2024}
}

@article{vfimamba,
  title={Vfimamba: Video frame interpolation with state space models},
  author={Zhang, Guozhen and Liu, Chunxu and Cui, Yutao and Zhao, Xiaotong and Ma, Kai and Wang, Limin},
  journal={Advances in Neural Information Processing Systems},
  volume={37},
  pages={107225--107248},
  year={2024}
}

@inproceedings{LBBDM,
  title={Frame interpolation with consecutive brownian bridge diffusion},
  author={Lyu, Zonglin and Li, Ming and Jiao, Jianbo and Chen, Chen},
  booktitle={Proceedings of the 32nd ACM International Conference on Multimedia},
  pages={3449--3458},
  year={2024}
}

@inproceedings{LDMVFI,
  title={Ldmvfi: Video frame interpolation with latent diffusion models},
  author={Danier, Duolikun and Zhang, Fan and Bull, David},
  booktitle={Proceedings of the AAAI Conference on Artificial Intelligence},
  volume={38},
  number={2},
  pages={1472--1480},
  year={2024}
}

@inproceedings{VIDIM,
  title={Video interpolation with diffusion models},
  author={Jain, Siddhant and Watson, Daniel and Tabellion, Eric and Poole, Ben and Kontkanen, Janne and others},
  booktitle={Proceedings of the IEEE/CVF Conference on Computer Vision and Pattern Recognition},
  pages={7341--7351},
  year={2024}
}

@inproceedings{dit,
  title={Scalable diffusion models with transformers},
  author={Peebles, William and Xie, Saining},
  booktitle={Proceedings of the IEEE/CVF International Conference on Computer Vision},
  pages={4195--4205},
  year={2023}
}

@inproceedings{pixart,
  title={PixArt-$\alpha $: Fast Training of Diffusion Transformer for Photorealistic Text-to-Image Synthesis},
  author={Chen, Junsong and Jincheng, YU and Chongjian, GE and Yao, Lewei and Xie, Enze and Wang, Zhongdao and Kwok, James and Luo, Ping and Lu, Huchuan and Li, Zhenguo},
  booktitle={The Twelfth International Conference on Learning Representations},
  year={2024}
}

@article{Latte,
  title={Latte: Latent diffusion transformer for video generation},
  author={Ma, Xin and Wang, Yaohui and Jia, Gengyun and Chen, Xinyuan and Liu, Ziwei and Li, Yuan-Fang and Chen, Cunjian and Qiao, Yu},
  journal={arXiv preprint arXiv:2401.03048},
  year={2024}
}

@inproceedings{ldm,
  title={High-resolution image synthesis with latent diffusion models},
  author={Rombach, Robin and Blattmann, Andreas and Lorenz, Dominik and Esser, Patrick and Ommer, Bj{\"o}rn},
  booktitle={Proceedings of the IEEE/CVF conference on computer vision and pattern recognition},
  pages={10684--10695},
  year={2022}
}

@inproceedings{sd3,
  title={Scaling rectified flow transformers for high-resolution image synthesis},
  author={Esser, Patrick and Kulal, Sumith and Blattmann, Andreas and Entezari, Rahim and M{\"u}ller, Jonas and Saini, Harry and Levi, Yam and Lorenz, Dominik and Sauer, Axel and Boesel, Frederic and others},
  booktitle={Forty-first International Conference on Machine Learning},
  year={2024}
}

@article{vae,
  title={Auto-encoding variational bayes},
  author={Kingma, Diederik P},
  journal={arXiv preprint arXiv:1312.6114},
  year={2013}
}

@inproceedings{magdiff,
  title={MagDiff: Multi-Alignment Diffusion for High-Fidelity Video Generation and Editing},
  author={Zhao, Haoyu and Lu, Tianyi and Gu, Jiaxi and Zhang, Xing and Zheng, Qingping and Wu, Zuxuan and Xu, Hang and Jiang, Yu-Gang},
  booktitle={European Conference on Computer Vision},
  pages={205--221},
  year={2025},
  organization={Springer}
}

@article{lstd,
  title={LSTD: Long Short-Term Temporal Diffusion for Video Generation},
  author={Zhao, Haoyu and Gu, Jiaxi and Wang, Shicong and Lu, Tianyi and Zhang, Xing and Wu, Zuxuan and Xu, Hang and Jiang, Yu-Gang},
  journal={IEEE Transactions on Multimedia},
  year={2026},
  publisher={IEEE}
}

@article{dynamictrl,
  title={Dynamictrl: Rethinking the basic structure and the role of text for high-quality human image animation},
  author={Zhao, Haoyu and Qi, Zhongang and Wang, Cong and Zheng, Qingping and Lu, Guansong and Chen, Fei and Xu, Hang and Wu, Zuxuan},
  journal={arXiv preprint arXiv:2503.21246},
  year={2025}
}

@inproceedings{cameranoise,
  title={CameraNoise: Enabling Faithful Camera Control in Video Diffusion through Geometry-Flow-Guided Noise Warping},
  author={Zhao, Haoyu and Gu, Jiaxi and Chen, Haoran and Zheng, Qingping and Jin, Yeying and Yang, Hongyi and Cheng, Junqi and Zhang, Yuang and Lu, Zenghui and Yu, Huan and others},
  booktitle={Forty-third International Conference on Machine Learning}
}

@article{svd,
  title={Stable video diffusion: Scaling latent video diffusion models to large datasets},
  author={Blattmann, Andreas and Dockhorn, Tim and Kulal, Sumith and Mendelevitch, Daniel and Kilian, Maciej and Lorenz, Dominik and Levi, Yam and English, Zion and Voleti, Vikram and Letts, Adam and others},
  journal={arXiv preprint arXiv:2311.15127},
  year={2023}
}

@article{sora,
  title={Sora: A review on background, technology, limitations, and opportunities of large vision models},
  author={Liu, Yixin and Zhang, Kai and Li, Yuan and Yan, Zhiling and Gao, Chujie and Chen, Ruoxi and Yuan, Zhengqing and Huang, Yue and Sun, Hanchi and Gao, Jianfeng and others},
  journal={arXiv preprint arXiv:2402.17177},
  year={2024}
}

@inproceedings{lpips,
  title={The unreasonable effectiveness of deep features as a perceptual metric},
  author={Zhang, Richard and Isola, Phillip and Efros, Alexei A and Shechtman, Eli and Wang, Oliver},
  booktitle={Proceedings of the IEEE conference on computer vision and pattern recognition},
  pages={586--595},
  year={2018}
}

@article{lavib,
  title={LAVIB: A Large-scale Video Interpolation Benchmark},
  author={Stergiou, Alexandros},
  journal={Advances in Neural Information Processing Systems},
  volume={37},
  pages={29091--29105},
  year={2024},
  address={Vancouver, Canada},
  month={December}
}

@article{davis,
  title={The 2017 davis challenge on video object segmentation},
  author={Pont-Tuset, Jordi and Perazzi, Federico and Caelles, Sergi and Arbel{\'a}ez, Pablo and Sorkine-Hornung, Alex and Van Gool, Luc},
  journal={arXiv preprint arXiv:1704.00675},
  year={2017}
}

@inproceedings{snufilm,
  title={Channel attention is all you need for video frame interpolation},
  author={Choi, Myungsub and Kim, Heewon and Han, Bohyung and Xu, Ning and Lee, Kyoung Mu},
  booktitle={Proceedings of the AAAI Conference on Artificial Intelligence},
  volume={34},
  number={07},
  pages={10663--10671},
  year={2020}
}

@inproceedings{stmfnet,
  title={St-mfnet: A spatio-temporal multi-flow network for frame interpolation},
  author={Danier, Duolikun and Zhang, Fan and Bull, David},
  booktitle={Proceedings of the IEEE/CVF Conference on Computer Vision and Pattern Recognition},
  pages={3521--3531},
  year={2022}
}

@article{motionfollower,
  title={Motionfollower: Editing video motion via lightweight score-guided diffusion},
  author={Tu, Shuyuan and Dai, Qi and Zhang, Zihao and Xie, Sicheng and Cheng, Zhi-Qi and Luo, Chong and Han, Xintong and Wu, Zuxuan and Jiang, Yu-Gang},
  journal={arXiv preprint arXiv:2405.20325},
  year={2024}
}

@article{ddpm,
  title={Denoising diffusion probabilistic models},
  author={Ho, Jonathan and Jain, Ajay and Abbeel, Pieter},
  journal={Advances in neural information processing systems},
  volume={33},
  pages={6840--6851},
  year={2020}
}

@inproceedings{flolpips,
  title={FloLPIPS: A bespoke video quality metric for frame interpolation},
  author={Danier, Duolikun and Zhang, Fan and Bull, David},
  booktitle={2022 Picture Coding Symposium (PCS)},
  pages={283--287},
  year={2022},
  organization={IEEE}
}

@inproceedings{hifi,
  title={High-resolution frame interpolation with patch-based cascaded diffusion},
  author={Hur, Junhwa and Herrmann, Charles and Saxena, Saurabh and Kontkanen, Janne and Lai, Wei-Sheng and Shih, Yichang and Rubinstein, Michael and Fleet, David J and Sun, Deqing},
  booktitle={Proceedings of the AAAI Conference on Artificial Intelligence},
  volume={39},
  number={4},
  pages={3868--3876},
  year={2025}
}

@inproceedings{eden,
  title={Eden: Enhanced diffusion for high-quality large-motion video frame interpolation},
  author={Zhang, Zihao and Chen, Haoran and Zhao, Haoyu and Lu, Guansong and Fu, Yanwei and Xu, Hang and Wu, Zuxuan},
  booktitle={Proceedings of the Computer Vision and Pattern Recognition Conference},
  pages={2105--2115},
  year={2025}
}

@inproceedings{tlbvfi,
  title={Tlb-vfi: Temporal-aware latent brownian bridge diffusion for video frame interpolation},
  author={Lyu, Zonglin and Chen, Chen},
  booktitle={Proceedings of the IEEE/CVF International Conference on Computer Vision},
  pages={16260--16269},
  year={2025}
}

@inproceedings{bimvfi,
  title={BiM-VFI: Bidirectional Motion Field-Guided Frame Interpolation for Video with Non-uniform Motions},
  author={Seo, Wonyong and Oh, Jihyong and Kim, Munchurl},
  booktitle={Proceedings of the Computer Vision and Pattern Recognition Conference},
  pages={7244--7253},
  year={2025}
}

@inproceedings{lcmamba,
  title={LC-Mamba: Local and continuous mamba with shifted windows for frame interpolation},
  author={Jeong, Min Wu and Rhee, Chae Eun},
  booktitle={Proceedings of the Computer Vision and Pattern Recognition Conference},
  pages={17671--17681},
  year={2025}
}

@article{pixeldit,
  title={PixelDiT: Pixel Diffusion Transformers for Image Generation},
  author={Yu, Yongsheng and Xiong, Wei and Nie, Weili and Sheng, Yichen and Liu, Shiqiu and Luo, Jiebo},
  journal={arXiv preprint arXiv:2511.20645},
  year={2025}
}

@article{pixnerd,
  title={Pixnerd: Pixel neural field diffusion},
  author={Wang, Shuai and Gao, Ziteng and Zhu, Chenhui and Huang, Weilin and Wang, Limin},
  journal={arXiv preprint arXiv:2507.23268},
  year={2025}
}

@article{dip,
  title={DiP: Taming Diffusion Models in Pixel Space},
  author={Chen, Zhennan and Zhu, Junwei and Chen, Xu and Zhang, Jiangning and Hu, Xiaobin and Zhao, Hanzhen and Wang, Chengjie and Yang, Jian and Tai, Ying},
  journal={arXiv preprint arXiv:2511.18822},
  year={2025}
}

@article{deco,
  title={Deco: Frequency-decoupled pixel diffusion for end-to-end image generation},
  author={Ma, Zehong and Wei, Longhui and Wang, Shuai and Zhang, Shiliang and Tian, Qi},
  journal={arXiv preprint arXiv:2511.19365},
  year={2025}
}

@article{jit,
  title={Back to basics: Let denoising generative models denoise},
  author={Li, Tianhong and He, Kaiming},
  journal={arXiv preprint arXiv:2511.13720},
  year={2025}
}

@article{wan,
  title={Wan: Open and advanced large-scale video generative models},
  author={Wan, Team and Wang, Ang and Ai, Baole and Wen, Bin and Mao, Chaojie and Xie, Chen-Wei and Chen, Di and Yu, Feiwu and Zhao, Haiming and Yang, Jianxiao and others},
  journal={arXiv preprint arXiv:2503.20314},
  year={2025}
}

@article{qwen-image,
  title={Qwen-image technical report},
  author={Wu, Chenfei and Li, Jiahao and Zhou, Jingren and Lin, Junyang and Gao, Kaiyuan and Yan, Kun and Yin, Sheng-ming and Bai, Shuai and Xu, Xiao and Chen, Yilei and others},
  journal={arXiv preprint arXiv:2508.02324},
  year={2025}
}

@article{oliva2006building,
  title={Building the gist of a scene: The role of global image features in recognition},
  author={Oliva, Aude and Torralba, Antonio},
  journal={Progress in brain research},
  volume={155},
  pages={23--36},
  year={2006},
  publisher={Elsevier}
}

@article{troje2008biological,
  title={Biological motion perception},
  author={Troje, Nikolaus F and Basbaum, A},
  journal={The senses: A comprehensive reference},
  volume={2},
  pages={231--238},
  year={2008}
}

@article{vaina2001functional,
  title={Functional neuroanatomy of biological motion perception in humans},
  author={Vaina, Lucia M and Solomon, Jeffrey and Chowdhury, Sanjida and Sinha, Pawan and Belliveau, John W},
  journal={Proceedings of the National Academy of Sciences},
  volume={98},
  number={20},
  pages={11656--11661},
  year={2001},
  publisher={The National Academy of Sciences}
}

@inproceedings{dmd,
  title={One-step diffusion with distribution matching distillation},
  author={Yin, Tianwei and Gharbi, Micha{\"e}l and Zhang, Richard and Shechtman, Eli and Durand, Fredo and Freeman, William T and Park, Taesung},
  booktitle={Proceedings of the IEEE/CVF conference on computer vision and pattern recognition},
  pages={6613--6623},
  year={2024}
}

@article{dmd2,
  title={Improved distribution matching distillation for fast image synthesis},
  author={Yin, Tianwei and Gharbi, Micha{\"e}l and Park, Taesung and Zhang, Richard and Shechtman, Eli and Durand, Fredo and Freeman, William T},
  journal={Advances in neural information processing systems},
  volume={37},
  pages={47455--47487},
  year={2024}
}

@inproceedings{add,
  title={Adversarial Diffusion Distillation},
  author={Sauer, Axel and Lorenz, Dominik and Blattmann, Andreas and Rombach, Robin},
  booktitle={European Conference on Computer Vision},
  pages={87--103},
  year={2024}
}

@inproceedings{mamba,
  title={Mamba: Linear-time sequence modeling with selective state spaces},
  author={Gu, Albert and Dao, Tri},
  booktitle={First conference on language modeling},
  year={2024}
}

\clearpage

\newpage

\subsection{Implementation Details}
\modelname~employs a base hidden state dimension of $d = 768$. To facilitate our macroscopic-to-microscopic generative curriculum, the progressive pixel-space architecture is partitioned into three distinct stages: 2 transformer blocks with a patch size of $P=64$, 6 blocks with $P=32$, and 4 blocks with $P=16$. During training, image instances are randomly cropped to spatial resolutions ranging from 384 to 1296 pixels to encourage robust scale invariance. The model is optimized over 100 epochs using the AdamW optimizer with a batch size of 64. The learning rate is modulated via a cosine annealing schedule, decaying smoothly from an initial value of $1 \times 10^{-4}$ down to $1 \times 10^{-8}$. Finally, the objective function weights are empirically set to $w_1 = 1.0$, $w_l = 1.0$, and $w_s = 20.0$ for the reconstruction, perceptual, and style losses, respectively.

\subsection{Pseudocode for One-Step Inference}
\label{sec:app_pseudocode}
To provide a concrete overview of our highly efficient inference pipeline, we present the pseudocode for the one-step generation process in Algorithm~\ref{alg:one_step_inference}. 

As a direct result of our Drift-aware Timestep Sampling (DTS) and the clean target ($x_0$) prediction objective, \modelname~completely bypasses the traditional iterative denoising loop. During inference, the framework requires only a single forward pass. By feeding the condition frames alongside a standard Gaussian noise tensor explicitly set to the maximum noise level ($t=1$), the model instantaneously projects the inputs into a high-quality intermediate frame in the pixel space.

\begin{algorithm}[h]
\caption{One-Step Inference of \modelname}
\label{alg:one_step_inference}
\begin{algorithmic}[1]
\Require Starting frame $I_0$, ending frame $I_1$, trained network $f_\theta$
\Ensure Interpolated intermediate frame $\tilde{I}_t$
\vspace{0.05in} % 微微增加一点间距，让输入输出与正文隔开，更透气

\Statex \(\triangleright\) Set timestep to maximum noise level
\State $t \leftarrow 1$

\Statex \(\triangleright\) Sample standard Gaussian noise matching frame dimensions
\State $\epsilon \sim \mathcal{N}(0, \mathbf{I})$

\Statex \(\triangleright\) Instantaneous forward projection in pixel space
\State $\tilde{I}_t \leftarrow f_\theta(\epsilon, I_0, I_1, t)$

\State \Return $\tilde{I}_t$
\end{algorithmic}
\end{algorithm}

\subsection{Visualization of Progressive Multi-Stage Representations}
\label{sec:app_visualization}
\begin{figure}[h]
    \centering
    \includegraphics[width=1.0\linewidth]{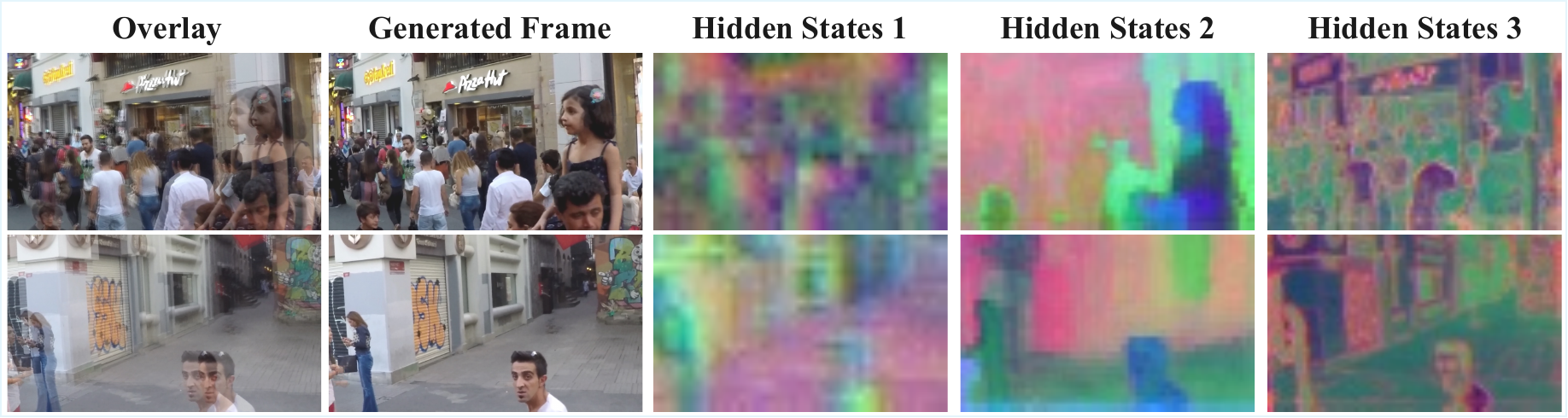}
    \caption{Visualization of the intermediate hidden states across the three stages of our progressive architecture.}
    \label{fig:stage_feats}
\end{figure}

To empirically validate the macroscopic-to-microscopic curriculum enforced by our progressive architecture, we visualize the intermediate hidden states extracted from the output of each stage. As illustrated in Fig.~\ref{fig:stage_feats}, the feature maps exhibit a clear, deliberate transition from coarse global motion to fine-grained local textures, effectively disentangling the optimization difficulty of pixel-space generation. 

\begin{itemize}
    \item \textbf{Stage 1: Macro-Motion Capture ($p$=64).} Aggressive tokenization yields coarse, abstract representations, forcing the network to establish global motion trajectories and large-scale structural layouts prior to focusing on local textures.
    \item \textbf{Stage 2: Structure \& Detail Refinement ($p$=32).} As patch size decreases, the hidden states resolve precise structural boundaries and continuous contours, effectively bridging coarse temporal dynamics with exact spatial geometry.
    \item \textbf{Stage 3: Micro-Texture Synthesis ($p$=16).} Retaining maximal spatial resolution, the tokens exhibit sharp, localized activations, indicating a complete computational shift toward synthesizing intricate micro-textures and fine appearance details.
\end{itemize}

This clear visual progression corroborates our architectural design, demonstrating that dynamic patch scaling successfully enforces a multi-scale generative curriculum.

\subsection{Influence of Drift-aware Timestep Sampling on Inference Steps}
\label{sec:app_dts_steps}
To further validate the effectiveness of our Drift-aware Timestep Sampling (DTS), we analyze its impact across varying numbers of inference steps. As detailed in the main paper, DTS dynamically collapses the sampling probability density toward the high-noise regime ($t=1$) during the later stages of training. Table~\ref{tab:ab_drift} demonstrates the critical necessity of this design for one-step generation.

\begin{table}[h]
\centering
\caption{Ablation study evaluating the influence of Drift-aware Timestep Sampling (DTS) across varying inference steps on DAVIS-256. The best results are highlighted in \textbf{bold}.}
% \resizebox{\columnwidth}{!}{
\begin{tabular}{cc cccc}
\toprule
\multirow{2}{*}{\textbf{Steps}} & \multirow{2}{*}{\textbf{Method}} & \multicolumn{4}{c}{\textbf{DAVIS-256}} \\
\cmidrule(lr){3-6}
 &  & PSNR$\uparrow$ & SSIM$\uparrow$ & LPIPS$\downarrow$ & FloLPIPS$\downarrow$ \\
\midrule
\multirow{2}{*}{1}
& w/o DTS & 28.7640 & 0.9032 & 0.0885 & 0.1268 \\
& \textbf{w/ DTS}   & \textbf{29.8569} & \textbf{0.9177} & \textbf{0.0639} & \textbf{0.0931} \\
\addlinespace
\multirow{2}{*}{2}
& w/o DTS & 29.2818 & 0.9090 & 0.0723 & 0.1062 \\
& \textbf{w/ DTS}   & \textbf{29.8575} & \textbf{0.9177} & \textbf{0.0638} & \textbf{0.0929} \\
\addlinespace
\multirow{2}{*}{5}
& w/o DTS & 29.3133 & 0.9094 & 0.0717 & 0.1055 \\
& \textbf{w/ DTS}   & \textbf{29.8581} & \textbf{0.9177} & \textbf{0.0638} & \textbf{0.0930} \\
\addlinespace
\multirow{2}{*}{10}
& w/o DTS & 29.3377 & 0.9099 & 0.0712 & 0.1048 \\
& \textbf{w/ DTS}   & \textbf{29.8550} & \textbf{0.9177} & \textbf{0.0638} & \textbf{0.0930} \\
\bottomrule
\end{tabular}
% }
\label{tab:ab_drift}
\end{table}

When attempting one-step inference without DTS, the model suffers a severe performance penalty, yielding an LPIPS of only 0.0885. Conversely, applying the DTS curriculum dramatically enhances one-step generative quality, improving the LPIPS to 0.0639. Notably, this instantaneous projection with DTS significantly outperforms even the 10-step iterative sampling of the baseline without DTS (LPIPS of 0.0712).

Furthermore, when DTS is active, increasing the number of inference steps beyond a single step yields negligible improvements across all evaluated metrics. The structural and perceptual quality at 1 step is nearly identical to the performance at 10 steps. This strongly confirms that our dynamic optimization curriculum successfully forces the network to completely internalize the one-step sampling distribution. Consequently, \modelname~achieves state-of-the-art quality while entirely bypassing the computational overhead of iterative sampling.

\subsection{Generalizability of One-Step VFI to Latent Diffusion Models}
\label{sec:app_latent_validation}
To demonstrate the generalizability of our proposed One-Step VFI, we extend our experiments to a state-of-the-art latent diffusion baseline, EDEN. We trained EDEN using our one-step learning framework for 100k steps at a spatial resolution of $144 \times 256$, evaluating the final one-step generation performance on the DAVIS-256 benchmark. The experiments, detailed in Table~\ref{tab:ab_predict_x_v_latent}, ablate the choice of the prediction target ($v$ vs. $x_0$) and the integration of our Drift-aware Timestep Sampling (DTS). 

Interestingly, while predicting the clean target ($x_0$) yields a massive performance boost in pixel space, the improvement in the latent space is relatively marginal. For instance, without DTS, shifting from predicting $v$ to $x_0$ only slightly improves the LPIPS from 0.1369 to 0.1345. This discrepancy arises from the mechanics of the VAE encoder: latent compression inherently entangles high-frequency details and low-frequency structures. Consequently, the dense structural correlation and high similarity between the intermediate frame and the boundary conditions—which are highly prominent in raw pixel space—are significantly degraded in the compressed latent space. Because the velocity field and the intermediate frames are not perfectly aligned across high and low dimensions in this compressed domain, the inherent advantage of direct $x_0$-prediction is weakened.

However, regardless of the prediction target, the application of DTS yields a substantial and consistent performance improvement. By dynamically shifting the training sampling density toward $t=1$, DTS successfully forces the latent network to internalize the instantaneous projection. When predicting $x_0$, introducing DTS improves the one-step LPIPS from 0.1345 down to 0.1149. This robust improvement demonstrates that while $x_0$-prediction is highly synergistic with pixel-space modeling, our DTS strategy is a universally effective optimization technique for enabling high-quality one-step generation in both pixel and latent diffusion frameworks.

\begin{table}[h]
\centering
\caption{Ablation study of one-step VFI evaluated on the latent diffusion baseline EDEN using DAVIS-256. DTS denotes our Drift-aware Timestep Sampling strategy.}
\begin{tabular}{cccccc}
\toprule
\multirow{2}{*}{\textbf{Model}} & \multirow{2}{*}{\textbf{predict\_v}} & \multirow{2}{*}{\textbf{predict\_x}} & \multirow{2}{*}{\textbf{DTS}} & \multicolumn{2}{c}{\textbf{DAVIS-256}} \\
\cmidrule(lr){5-6}
 & & & & LPIPS$\downarrow$ & FloLPIPS$\downarrow$ \\ 
\midrule
\multirow{4}{*}{EDEN} & \Checkmark &   \XSolidBrush     & \XSolidBrush & 0.1369             & 0.1875             \\
                      & \Checkmark &   \XSolidBrush     & \Checkmark  & \underline{0.1165} & \underline{0.1563} \\
\cmidrule(lr){2-6}
                      &    \XSolidBrush     & \Checkmark & \XSolidBrush & 0.1345 & 0.1868 \\
                      &   \XSolidBrush      & \Checkmark & \Checkmark  & \textbf{0.1149}    & \textbf{0.1541}    \\ 
\bottomrule
\end{tabular}
\label{tab:ab_predict_x_v_latent}
\end{table}

\subsection{More Visualization Results}
\label{sec:app_more_vis}
\begin{figure*}[htbp]
\centering
\includegraphics[width=\textwidth]{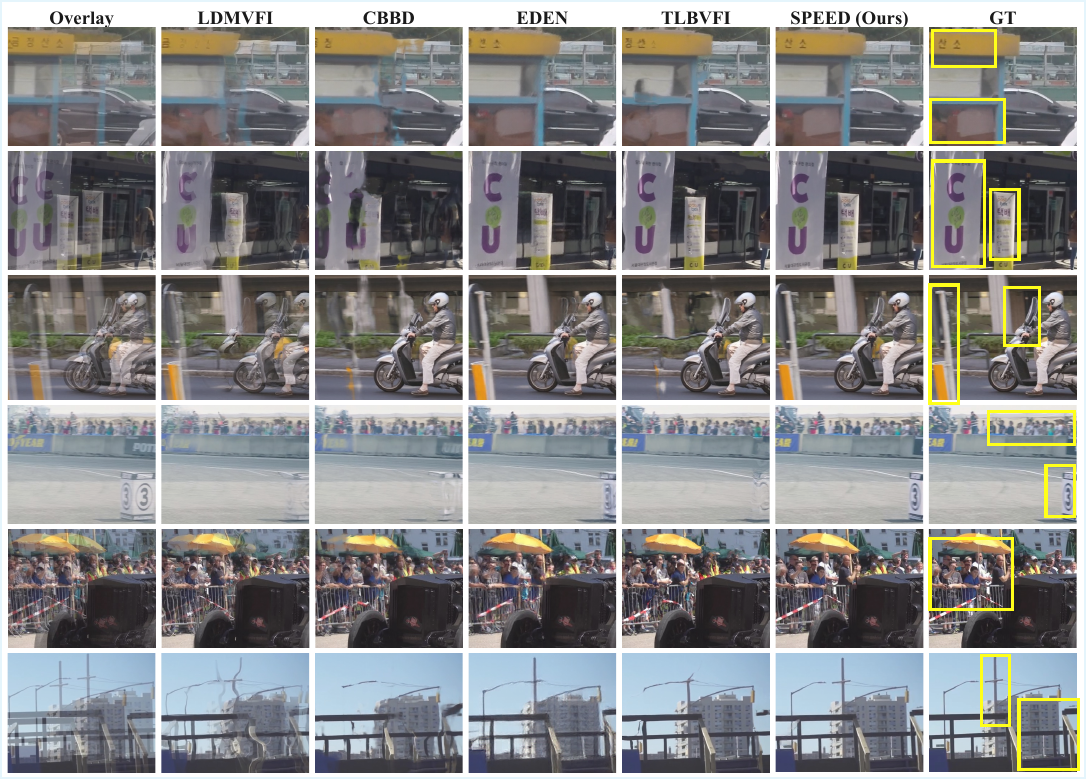}
\caption{Visual comparison of \modelname~against state-of-the-art diffusion-based methods across challenging large-motion and complex textures scenarios. By operating in pixel space, \modelname~consistently outperforms previous latent diffusion baselines, preserving structural integrity and sharp micro-textures without VAE-induced blurring or ghosting.}
\label{fig:appendix_visualization}
\end{figure*}

To further demonstrate the superiority of our proposed \modelname~framework, we provide extensive visual comparisons against state-of-the-art latent diffusion baselines (LDMVFI, CBBD, EDEN, and TLB-VFI) in Figure~\ref{fig:appendix_visualization}. The selected scenarios feature highly challenging non-linear motions, complex textures, and delicate structural elements.

As clearly observed, methods relying on latent compression consistently suffer from VAE-induced artifacts. This limitation manifests as severe blurring of textual elements, ghosting around fast-moving subjects (e.g., the scooter and vehicles), and the outright destruction of high-frequency thin structures (e.g., background wires and poles). In contrast, by operating directly within the continuous pixel space, \modelname~entirely circumvents these latent compression bottlenecks. Our approach consistently preserves sharp micro-textures, maintains precise structural boundaries during large-scale motions, and reconstructs fine-grained spatial details that closely align with the Ground Truth (GT).

\subsection{Multi-frame Interpolation}
To evaluate the temporal consistency and robustness of \modelname~in extreme large-motion scenarios, we conduct an $8\times$ multi-frame interpolation experiment. By recursively applying our one-step generation process, we synthesize 7 intermediate frames between a given starting and ending frame. 

\begin{figure*}[htbp]
\centering
\includegraphics[width=\textwidth]{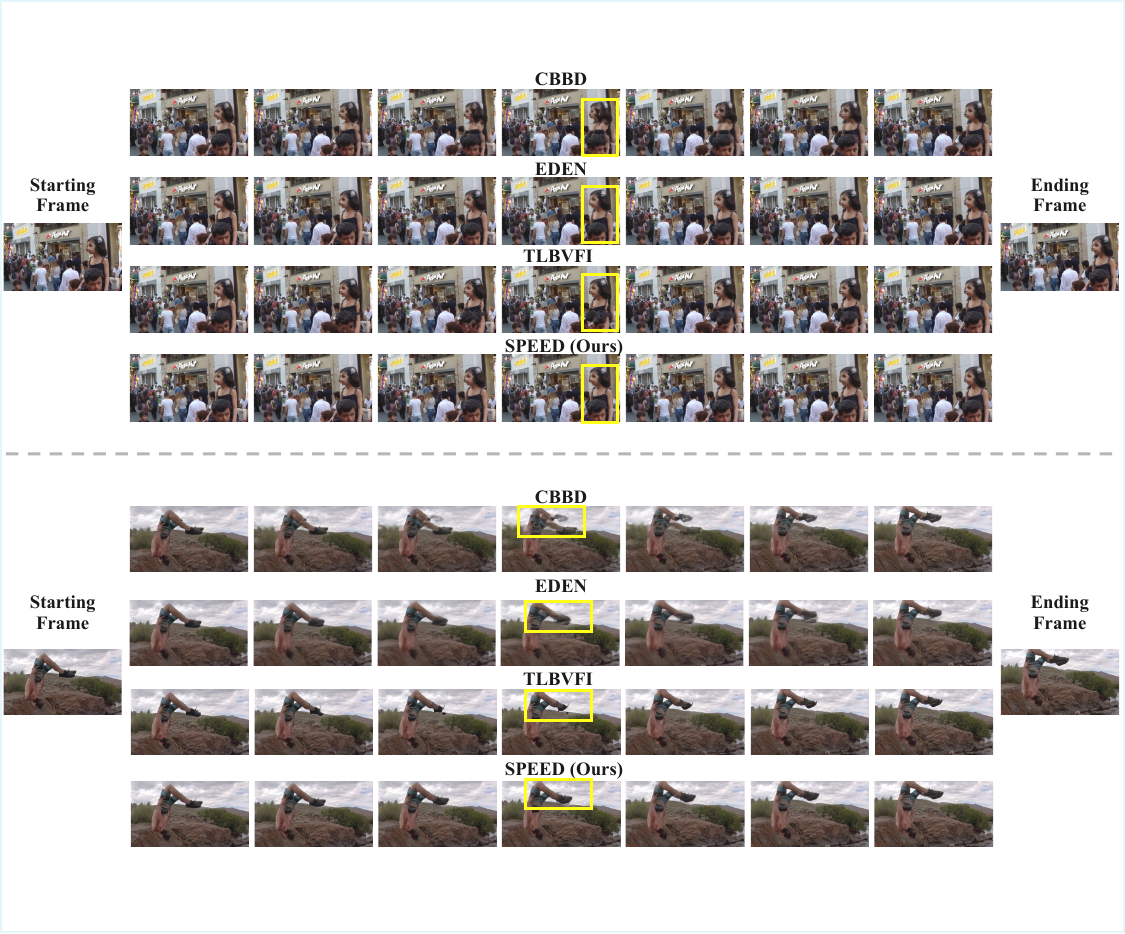}
\caption{Visual comparison of $8\times$ multi-frame interpolation. \modelname~consistently outperforms previous latent diffusion baselines, maintaining superior structural integrity and temporal coherence across all 7 generated intermediate frames without compounding artifact accumulation.}
\label{fig:multi_frames_visualization}
\end{figure*}

As illustrated in Figure~\ref{fig:multi_frames_visualization}, baseline latent diffusion models (CBBD, EDEN, and TLB-VFI) struggle significantly with recursive generation. They exhibit severe error accumulation, temporal jitter, and structural degradation as the sequence progresses. In contrast, \modelname~maintains remarkable structural integrity, sharp micro-textures, and smooth motion trajectories across the entire sequence. This confirms that our direct pixel-space modeling and optimized single-step projection provide highly stable spatio-temporal priors, effectively mitigating the compounding artifacts that commonly plague recursive latent diffusion.

\end{document}